\documentclass[lettersize,journal]{IEEEtran}
\usepackage{graphicx} 
\usepackage{enumerate}
\usepackage{amsmath,amsfonts,amssymb,amsthm}
\usepackage{algorithm,algorithmic}
\usepackage{comment}
\usepackage[caption=false,font=normalsize,labelfont=sf,textfont=sf]{subfig}
\usepackage{textcomp}
\usepackage{float}
\usepackage{stfloats}
\newtheorem{theorem}{Theorem}
\newtheorem{proposition}{Proposition}

\newtheorem{lemma}{Lemma}

\usepackage{mathtools}
\usepackage{multicol}
\usepackage{xcolor}
\usepackage{xspace}
\usepackage{multirow}
\usepackage{cite}
\usepackage{comment}
\usepackage{booktabs}

\title{Handover Cost Optimization in LEO Satellite Networks using an RL-based Multiple-Column Selection Strategy via Branch-and-Price}
\author{Mandar R. Nalavade, Gaurav S. Kasbekar,~\IEEEmembership{Senior Member,~IEEE}, Koushik Kar,~\IEEEmembership{Senior Member,~IEEE}

\thanks{The authors Mandar R. Nalavade, Gaurav S. Kasbekar are with the Department of Electrical Engineering, Indian Institute of Technology (IIT) Bombay, Mumbai, 400076, India and the author Koushik Kar is with the Department of Electrical, Computer \& Systems Engineering, Rensselaer Polytechnic Institute New York, United States. Their email addresses are 22d0531@iitb.ac.in, gskasbekar@ee.iitb.ac.in, and koushik@ecse.rpi.edu. 
}}
\date{}

\begin{document}

\maketitle

\begin{abstract}  
    Next-generation wireless networks increasingly rely on low Earth orbit (LEO) satellites for broad coverage and low latency, making efficient handover management essential. We formulate satellite handover as a combinatorial optimization problem, assigning each ground user to exactly one capacity-limited, visible satellite while minimizing network handover costs. We propose a reinforcement learning (RL)-based branch-and-price (B\&P) framework with learned multiple-column selection (MCS) to accelerate column generation while retaining exact dynamic programming pricing. We also develop two lightweight heuristic policies: handover-aware visibility duration (HAVD) for online decisions and congestion-aware predictive (CAP) for offline planning. Additionally, a local-search repair algorithm adapts precomputed B\&P or MCS-B\&P solutions to demand perturbations through single-flip and per-user trajectory re-optimization, avoiding complete re-optimization. Simulations using realistic large scale constellations, including Starlink Gen-1, demonstrate that MCS-B\&P achieves solution quality comparable to exact B\&P with fewer column-generation iterations, fewer branch-and-bound nodes, and reduced computation time. HAVD and CAP offer effective online and offline cost-latency trade-offs, while local-search repair recovers most performance losses due to demand perturbations at a fraction of the computational cost of full re-optimization.
\end{abstract}


\section{Introduction} \label{sec:introduction}
Satellite communications play an important role in providing seamless global connectivity and overcoming the inherent limitations of terrestrial mobile communications. Low Earth orbit (LEO) constellations, with orbits deployed at altitudes between $500$ km and $1500$ km, allow lower transmission latency and higher throughput compared to geostationary Earth orbit (GEO) satellites \cite{darwish2022leo}. In recent years, the demand for broadband services offered by large constellations of LEO satellite networks (e.g., Kuiper, Starlink, OneWeb, and Lightspeed) has increased. For example, Starlink provides low latency (below $30$ ms), high data rate ($\geq 100$ Mb/s) satellite broadband services to inaccessible areas \cite{araniti2021toward, zhou2023aerospace}. As constellation sizes continue to grow, efficient handover management has become a critical challenge for next generation satellite communication systems. Unlike traditional geostationary systems, LEO satellites move rapidly relative to ground users, and complete an orbit around the Earth in about $90-120$ minutes \cite{tuysuz2006satellite}. Therefore, each satellite can serve the gateway stations or ground users for a limited time, which is usually shorter than the duration of a connection \cite{wang2021deep}. To ensure uninterrupted communication, once the current satellite is no longer able to provide services to the gateway station, it is necessary to handover the communication link to another visible satellite. The rapid motion of LEO satellites causes constant changes in constellation topology, resulting in frequent handovers, and significantly affects quality of service (QoS).

A satellite handover optimization problem is combinatorial in nature. Early approaches primarily relied on myopic or elevation-based selection strategies, where users connect to the satellite with the strongest instantaneous link quality or highest elevation angle. While computationally simple, these methods often lead to frequent and unnecessary handovers, resulting in increased signaling overhead and reduced service stability \cite{eki2001distributed}. To mitigate these issues, optimization-based formulations have been proposed that jointly consider handover cost, resource constraints, and service continuity over a finite horizon. In particular, time-indexed integer programming models and dynamic programming (DP) approaches have been developed to capture the temporal coupling of handover decisions, although their scalability remains limited for large constellations and long time horizons \cite{kod2021satellite}. More recently, graph-based formulations \cite{hozayen2023graph}, learning-assisted techniques \cite{wang2021deep, yu2024graph, he2020load, lang2026joint, zhang2026jointly, sun2024handover, ntabeni2025adaptive} have been explored to improve computational efficiency and adaptability in dynamic satellite environments. Despite these advances, achieving scalability while retaining optimality or near-optimality for large-scale satellite handover optimization remains challenging, motivating the development of complementary optimization, learning-assisted, and lightweight decision-making approaches.

The branch-and-price (B\&P) framework has emerged as a powerful exact solution methodology for large scale combinatorial optimization problems characterized by an exponential number of decision variables \cite{barnhart1998branch}. By integrating column generation (CG) within a branch and bound (B\&B) tree, B\&P enables the implicit handling of huge variable spaces while preserving global optimality guarantees. The framework has been successfully applied across a wide range of domains, including vehicle routing, crew scheduling, cutting stock, network design, and telecommunications resource allocation problems \cite{desrosiers2024branch}. In such applications, the master problem typically selects among complex structures such as routes, schedules, or paths, while the pricing subproblem generates new columns by solving auxiliary optimization problems, often formulated as shortest-path or DP models. The effectiveness of B\&P largely stems from the strong linear programming (LP) relaxations obtained through problem specific decompositions, which significantly reduce the size of the B\&B tree. Despite its success, classical B\&P methods can face computational bottlenecks due to the repeated solution of pricing problems, motivating the development of more efficient column-generation and pricing strategies, including learning-assisted approaches. Comprehensive treatments of the theory and applications of B\&P can be found in the works of \cite{barnhart1998branch, desrosiers2024branch, desrosiers2005primer}. The work in \cite{jiang2022branch, pei2025branch, liang2024optimizing} uses B\&P for optimization purposes.

Recent research has explored the integration of reinforcement learning (RL) with exact optimization frameworks such as B\&P to address the computational challenges posed by large scale combinatorial problems. RL-assisted B\&P approaches aim to mitigate these bottlenecks by learning policies that guide CG) branching decisions, or node selection, thereby accelerating convergence without sacrificing optimality. In particular, RL can be used to efficiently explore large structured decision spaces and propose high quality candidate columns, while exact reduced cost evaluation ensures correctness of the master problem solution. Such hybrid learning optimization paradigms have shown promising results in vehicle routing \cite{furian2021machine, abouelrous2025reinforcement} and scheduling \cite{morabit2023machine, morabit2021machine}, where RL improves search efficiency and reduces computational overhead compared to purely exact methods. Recently, machine learning (ML) has also been employed in the CG procedure to enhance the performance of exact optimization methods \cite{vaclavik2018accelerating, khalil2016learning, yuan2024reinforcement, shen2022enhancing, chi2022deep}. Designing RL components that speed up B\&P while still maintaining its optimality guarantees remains a challenging and ongoing research problem. Several recent studies have explored the use of RL to enhance the performance of the classical B\&P framework \cite{furian2021machine, abouelrous2025reinforcement, morabit2023machine, morabit2021machine, vaclavik2018accelerating} and exact methods \cite{khalil2016learning, yuan2024reinforcement}. However, to the best of our knowledge, an RL-based B\&P specifically tailored for the satellite handover optimization problem has not been investigated in the existing literature \cite{furian2021machine, abouelrous2025reinforcement, morabit2023machine, morabit2021machine, vaclavik2018accelerating, khalil2016learning, yuan2024reinforcement}.

The work \cite{yuan2024reinforcement}, introduced an RL-based multiple-column selection (MCS) strategy for general CG problems: instead of adding the single best column each iteration, a learned policy selects $k$ columns from an $n$-candidate pool, achieving faster LP convergence on cutting-stock and graph-coloring benchmarks. We adapt and extend this framework to the satellite handover problem, making three principal contributions: 
\begin{enumerate}
    \item Formulation of the handover CG as a Markov decision process (MDP) whose state captures the bipartite restricted master problem (RMP) graph, candidate trajectory features, and dual-variable information, with a reward primarily driven by LP objective improvement and augmented by candidate diversity and dual progress;
    \item Design of an actor--critic architecture comprising a graph isomorphism network (GIN) based bipartite encoder, a graph attention network (GAT) based candidate--interaction decoder, and a global feature branch, and pre-train the policy offline on the target problem instance to adapt the column-selection strategy to the underlying constellation geometry, visibility structure, and demand distribution;
    \item Integration of the MCS policy into a full B\&P solver (MCS-B\&P) with a tuned per-node CG budget, diversity aware initial column pool, and perturbation based pricing runs that promote structurally diverse trajectories. On Starlink scenarios, MCS-B\&P achieves improved wall-clock performance compared with classical B\&P while maintaining solution quality.
\end{enumerate}
Apart from this, following points are the major contributions regarding the complementary handover aware visibility duration (HAVD) policy, and the congestion aware predictive (CAP) policy:
\begin{enumerate}
    \item A capacity-gated, weighted-scoring online heuristic, termed HAVD, together with a formal complexity analysis;
    \item An offline capacity-aware predictive strategy, termed CAP, that computes the service duration and minimum congestion slack (MS) for each candidate satellite along with complexity analysis;
    \item A comprehensive comparison of HAVD and CAP on a Starlink Gen-1 constellation, showing the trade-off between handover cost and computational time and characterizing the performance gap between the two policies.
\end{enumerate}
Beyond nominal-demand performance, practical LEO deployments must also cope with fluctuations in user traffic that are not known precisely at planning time. To address this, we study the robustness of the B\&P solution under demand perturbation and introduce a local-search repair algorithm that adapts a precomputed B\&P assignment to perturbed demand without re-solving the B\&P problem from scratch. The algorithm combines a single-flip neighborhood, which locally reassigns individual user-time slot satellite links to restore capacity feasibility and reduce handover cost, with a per-user trajectory re-optimization that re-solves a single user's entire trajectory exactly while holding all other assignments fixed, enabling the search to escape local optima that single-flip moves alone cannot reach. We evaluate this repair procedure across a range of demand perturbation magnitudes and show that it recovers most of the cost degradation induced by the perturbation at a small fraction of the computational cost of an exact B\&P re-solve, while we also quantify the residual optimality gap relative to the exact re-solved solution.

The remainder of this paper is organized as follows. Section \ref{section_related_work} reviews the relevant prior work. Section \ref{section_motivation} provides the motivation behind the work. In Section \ref{SC:system:model}, we present the system model and formulate the problem. Section \ref{methodology} provides the exact B\&P framework methodology in detail. Our proposed RL-based MCS-B\&P is explained in Section \ref{true_RL}. Sections \ref{online_policy} and \ref{offline_policy} describe online and offline handover policies respectively. Computationally fast local-search repair algorithm under demand perturbation is described in Section \ref{sec:local_search_repair}. The simulation results and comparison of the proposed algorithm with the classical B\&P, offline and online handover policies are provided in Section \ref{results}. Finally, Section \ref{conclusion} concludes and outlines potential avenues for future research.

\section{Related Work} \label{section_related_work}
Several optimization-based and heuristic methods have been proposed to address the satellite handover problem in LEO networks.
A comprehensive survey of handover management schemes in LEO satellite networks was presented in \cite{chowdhury2006handover}, systematically categorizing them into link-layer and network-layer approaches and evaluating their performance based on various QoS criteria.
The satellite handover problem in LEO networks was analyzed in \cite{papapetrou2004satellite, papapetrou2005analytic}, which introduced an optimization-based dynamic Doppler-based handover prioritization technique to enhance user QoS.
The study in \cite{voicu2024handover} analyzed the impact of inter-satellite handover (HO) strategies on space-to-ground link performance, considering three approaches: closest-satellite handover, maximum-visibility handover, and carrier to interference plus noise ratio (CINR)-based handover triggered when the CINR falls below a threshold. Their impact was also evaluated across LEO, medium Earth orbit (MEO), and highly elliptical orbit (HEO) satellite networks.
In \cite{wang2024satellite}, the authors studied handover management for multiple civil aircraft in ultra-dense LEO satellite networks (UDLSNs), considering aircraft priorities and a satisfaction function incorporating rate satisfaction, power overhead, and handover overhead. They formulated the problem as a local cooperative game, established the existence of a Nash equilibrium (NE), and proposed a parallel handover strategy update (PHSU) algorithm to achieve it.
A graph-based framework to model and analyze the satellite handover process in LEO satellite communication networks was introduced and assessed in \cite{wu2016graph}. However, none of the above works \cite{chowdhury2006handover, papapetrou2004satellite, papapetrou2005analytic, voicu2024handover, wang2024satellite, wu2016graph} addresses the satellite handover optimization problem using the RL-based multiple-column selection B\&P (MCS-B\&P) framework. 

ML-based approaches have recently gained significant attention for satellite handover optimization in dynamic LEO environments.
A novel deep reinforcement learning (DRL)-based handover scheme for LEO networks was proposed in \cite{wang2021deep}, accounting for multiple handover factors simultaneously. 
In \cite{yu2024graph}, a graph reinforcement learning-based handover strategy combining graph neural networks and deep Q-learning was proposed to optimize path selection in dynamic LEO networks while accounting for network topology and multiple QoS factors.
A multi agent RL–based strategy for LEO satellite networks was proposed in \cite{he2020load}, aiming to minimize average satellite handovers while satisfying the load constraints of each satellite.
In \cite{lang2026joint}, an age of information (AoI)-aware space-air-ground integrated network (SAGIN) architecture employing a high-altitude platform as a relay was proposed to jointly minimize AoI and satellite handover frequency using a diffusion-enhanced dueling double deep Q-network (DD3QN).
A dynamic dual agent optimization framework (D2AOF) was proposed in \cite{zhang2026jointly} to jointly optimize handovers and power allocation in LEO satellite networks, integrating deep Q-networks (DQN) for handover decisions and proximal policy optimization (PPO) for power allocation within an iterative intelligent framework.
In \cite{sun2024handover}, a multi-objective deep Q-network (MODQN)-based beam handover strategy was proposed for multi-beam LEO satellite networks to maximize throughput while minimizing handover frequency.
A Q-learning based energy-aware handover framework for a single user remote scenarios was studied in \cite{ntabeni2025adaptive}, focusing on improving signal quality while reducing unnecessary handovers and their associated energy costs.
However, none of the above works \cite{wang2021deep, yu2024graph, he2020load, lang2026joint, zhang2026jointly, sun2024handover, ntabeni2025adaptive} addresses the satellite handover optimization problem in LEO networks using the RL-based MCS-B\&P framework. 

Exact optimization approaches, including B\&P, have demonstrated strong performance in solving large scale combinatorial problems.
In \cite{jiang2022branch}, the authors studied a large scale multi-depot electric bus scheduling problem under vehicle–depot constraints and partial recharging. They formulated it as a mixed integer program and solved it using an efficient B\&P framework.
The urban aerial delivery problem (UADP) was formulated as a mixed-integer linear program in \cite{pei2025branch}, where a set covering formulation with linear relaxation was introduced to improve computational efficiency. A B\&P algorithm incorporating heuristic pricing acceleration was then developed to solve the model efficiently.
The joint deployment of fixed and vehicle-based roadside units (RSUs) was formulated as a mixed-integer optimization problem in \cite{liang2024optimizing}, solved using CPLEX after linearization. The authors also developed a route-based formulation and a B\&P approach with a heuristic for initial solution generation.
In \cite{danach2025multi}, a multi-depot capacitated vehicle routing problem with time windows and multi-modal transportation was formulated to jointly optimize hub locations and delivery routes using autonomous delivery vehicles (ADVs) and drones. The problem was solved using a B\&P framework strengthened with valid inequalities.
A multi-objective mixed-integer program for home care planning was developed in \cite{alkaabneh2023multi}, minimizing service and routing costs while maximizing nurse–patient compatibility. The authors proposed a B\&P framework and a two stage metaheuristic to solve the problem. However, none of the above works \cite{jiang2022branch, pei2025branch, liang2024optimizing, danach2025multi, alkaabneh2023multi} utilizes B\&P technique to solve the satellite handover optimization problem.

Recent studies have investigated the use of ML techniques to accelerate the CG procedure in exact optimization frameworks.
An ML approach was introduced in \cite{khalil2016learning}, to learn branching variable rankings from strong branching decisions, reducing B\&B search tree size with low computational overhead.
An RL-based multiple-column selection strategy for CG, modeling the selection process as a sequential decision problem using an actor–critic framework was proposed in \cite{yuan2024reinforcement}. The approach learned to choose the best combination of columns by considering their interactions and global problem structure.
An ML-based pricing heuristic was proposed in \cite{shen2022enhancing} to accelerate CG by predicting and sampling solutions to NP-hard pricing problems, generating high-quality columns for graph coloring and improving convergence while reducing computational time.
A DRL framework was proposed in \cite{chi2022deep} to guide column selection in CG using a graph neural network, optimizing long-term rewards to reduce iterations and runtime compared with greedy decisions. However, none of the above works \cite{khalil2016learning, yuan2024reinforcement, shen2022enhancing, chi2022deep} addresses the problem of satellite handover optimization problem in LEO networks.

The integration of ML techniques within B\&P frameworks has recently emerged as a promising direction for solving large scale combinatorial optimization problems.
An ML-based B\&P framework was proposed in \cite{furian2021machine} for structured vehicle routing problems with time windows and capacity constraints, using ML models to predict optimal binary variables and estimate branching scores for fractional variables.
An RL model based on an attention-mechanism architecture was studied in \cite{abouelrous2025reinforcement} to identify columns with the most negative reduced cost in the pricing problem, with a vehicle routing problem considered as the case study.
An ML-based pricing heuristic was proposed in \cite{morabit2023machine} to accelerate pricing by retaining arcs likely to appear in the linear relaxation, and was evaluated on public transit vehicle and crew scheduling and vehicle routing with time windows.
A column-selection heuristic was proposed in \cite{morabit2021machine} to accelerate CG by selecting promising columns to reduce RMP re-optimization time, with effectiveness demonstrated on vehicle and crew scheduling and vehicle routing with time windows.
In \cite{vaclavik2018accelerating}, an ML-based method was introduced that extends the generic B\&P framework, rather than being problem specific. It incorporated a regression model to retain the learned knowledge and to predict upper bounds (UBs) using problem specific features.
However, none of the above works \cite{abouelrous2025reinforcement, furian2021machine, vaclavik2018accelerating, morabit2023machine, morabit2021machine} uses the RL-based MCS-B\&P framework to solve the satellite handover optimization problem in LEO networks.

Local-search and heuristic approaches have been widely investigated as computationally efficient alternatives for large-scale network optimization. 
In \cite{di2024deepls}, a local-search framework is developed for network optimization in which local moves are used to improve solutions for NP-hard routing problems, demonstrating competitive solution quality with substantially reduced computation time. 
Similarly, the work \cite{van2020distributed} considered distributed wireless network optimization and employed stochastic local search to handle large optimization spaces while limiting communication and computational requirements. 
A work in \cite{zhang2022satellite} proposed a heuristic handover strategy that jointly considers user experience, satellite throughput, and network load, with the heuristic search introduced to reduce the computational cost of handover optimization. 
In \cite{hozayen2023graph} a graph-based handover framework for LEO satellite networks formulated handover timing and target selection as a shortest-path problem and reported low computational complexity on a Starlink Phase-I constellation. 
The work \cite{chen2024shortest} proposed STEPCLIMB, an explicit phase-based algorithm that exploited the regular topology of LEO constellations to solve the shortest-path problem without iterative graph search. The method achieved low computational cost while maintaining high path accuracy in large-scale LEO constellations. 
A weighted bipartite-graph-based handover strategy for LEO satellite networks is proposed in \cite{eydian2025handover}, using maximum-weight matching to balance link quality and satellite load. They further incorporated a hysteresis margin to reduce unnecessary handovers and improve handover rate, latency, and overall QoS. 
A hierarchical RL framework that separately learned when to initiate a handover and which satellite to select is designed in \cite{zhao2026learning}, considering load, signal quality, and service duration. The method improved handover frequency, success rate, transmission rate, and load balancing in a Starlink environment. However, none of the above works \cite{di2024deepls, van2020distributed, zhang2022satellite, hozayen2023graph, chen2024shortest, eydian2025handover, zhao2026learning} explicitly address capacity-aware online and offline handover decisions or repair of an existing solution after demand perturbations.

\section{Motivation} \label{section_motivation}
The dynamic nature of LEO satellite networks introduces significant challenges in maintaining stable and efficient user–satellite associations. Due to the rapid motion of satellites, both the set of visible satellites and the link quality experienced by users evolve continuously over time, making handover decisions inherently time coupled and complex.

\begin{figure}[ht]
\centering
\includegraphics[width=0.48\textwidth]{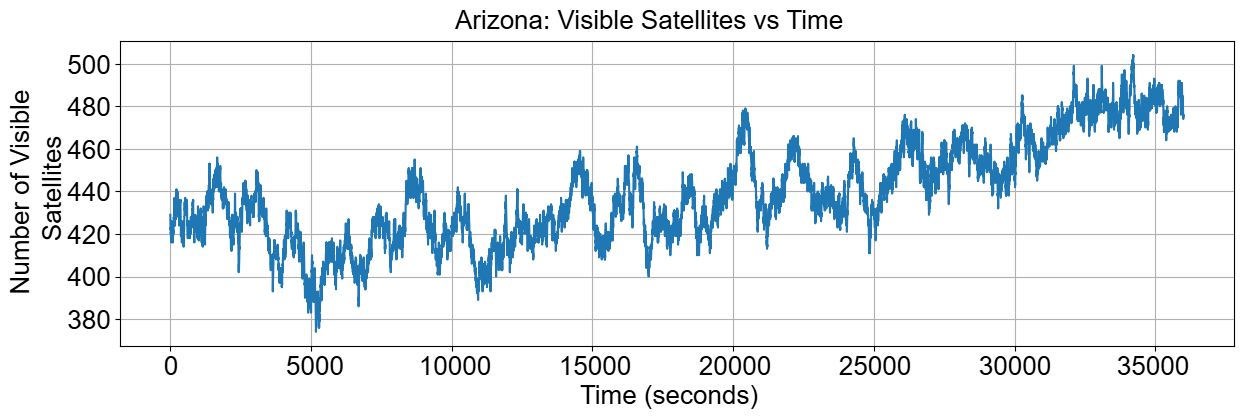}
\caption{The figure shows the number of visible STARLINK satellites from Arizona over $12$ hours.}
\label{motiv:fig:1}
\end{figure}

\begin{figure}[ht]
\centering
\includegraphics[width=0.48\textwidth]{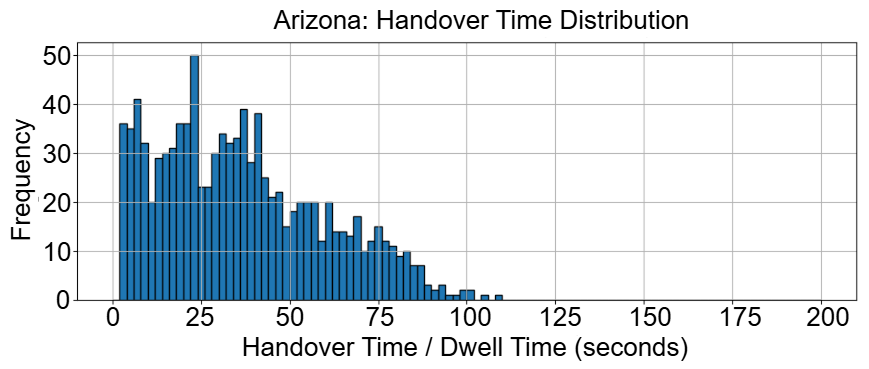}
\caption{The figure shows the histogram of handover time within a span of $2$ seconds over Arizona.}
\label{motiv:fig:2}
\end{figure}

\begin{figure}[ht]
\centering
\includegraphics[width=0.48\textwidth]{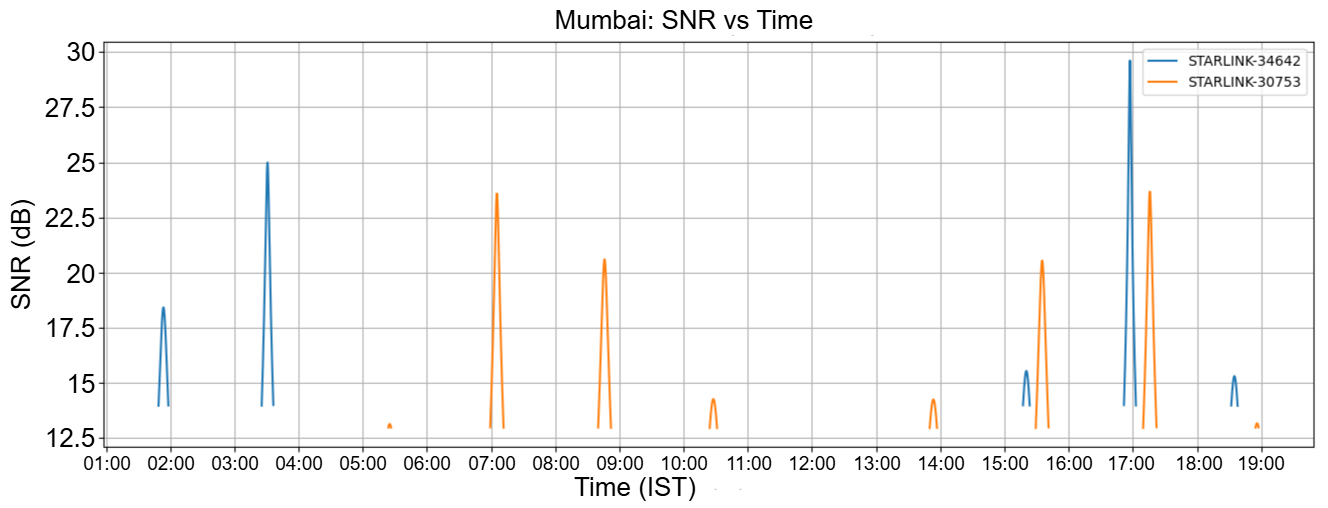}
\caption{The figure shows SNR variation over time for selected satellites in Mumbai over a day.}
\label{motiv:fig:3}
\end{figure}

\begin{figure}[ht]
\centering
\includegraphics[width=0.48\textwidth]{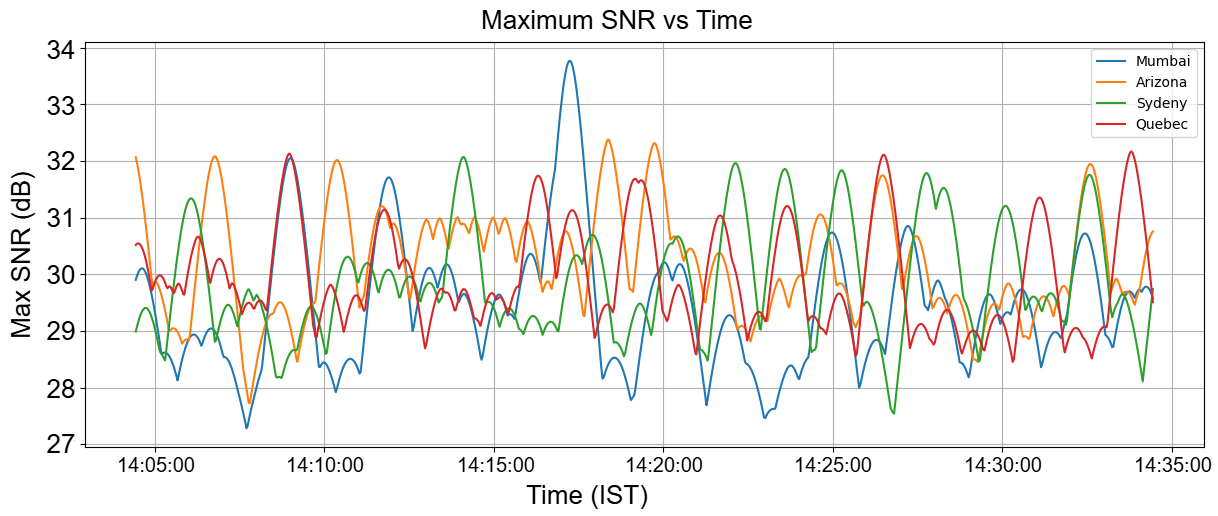}
\caption{The figure shows Maximum SNR variation over a specific time across different cities.}
\label{motiv:fig:4}
\end{figure}

\begin{figure}[ht]
\centering
\includegraphics[width=0.48\textwidth]{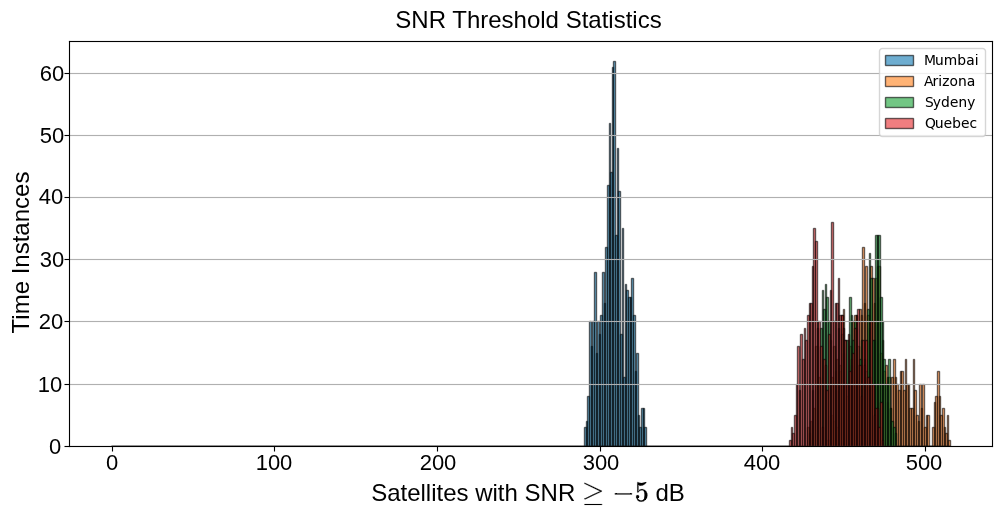}
\caption{The figure shows the histogram of satellites meeting the SNR threshold $(\geq -5 dB)$ across multiple cities.}
\label{motiv:fig:5}
\end{figure}

\begin{figure}[ht]
\centering
\includegraphics[width=0.48\textwidth]{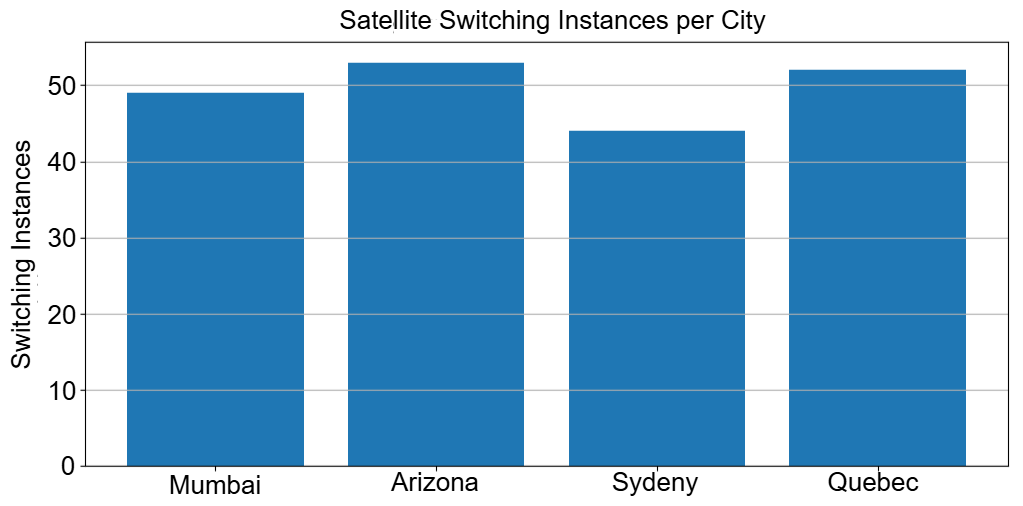}
\caption{The figure shows the number of satellite handover (switching) instances observed across different cities.}
\label{motiv:fig:6}
\end{figure}

To better understand these challenges, Figs. \ref{motiv:fig:1}, \ref{motiv:fig:2}, \ref{motiv:fig:3}, \ref{motiv:fig:4}, \ref{motiv:fig:5}, and \ref{motiv:fig:6} in this section are generated using a simulation framework based on real Starlink satellite orbital data obtained from publicly available two-line element (TLE) datasets. The satellite positions are computed using the Skyfield library, and link-level signal-to-noise ratio (SNR) is evaluated using a standard satellite communication link budget model using the parameters provided in Table \ref{tab:parameter_values}. Ground stations corresponding to multiple geographic locations (e.g., Mumbai, Arizona, Sydney, and Quebec) are simulated over a finite time horizon. At each time step, the visible satellites are identified based on the elevation-angle constraint, and the resulting number of visible satellites is recorded along with the maximum SNR and number of handover instances for performance evaluation.

At first, the temporal variation in satellite visibility is highly pronounced. As shown in Fig. \ref{motiv:fig:1}, which illustrates the number of visible satellites over time for the Arizona region, the visibility fluctuates significantly with both short term oscillations and long term trends. This indicates that the candidate satellite set for each user is continuously changing, and decisions based solely on instantaneous conditions may lead to unstable associations. The handover (dwell) time characteristics highlight the transient nature of user–satellite links. As shown in Fig. \ref{motiv:fig:2}, most connections last only a few tens of seconds, with a long tail of longer durations. This skewed distribution leads to frequent and often unnecessary handovers under myopic strategies, increasing signaling overhead, degrading QoS, and reducing overall network efficiency, now supported by empirical evidence.

The signal quality dynamics further complicate the decision making process. The SNR variation over time for Mumbai, shown in Fig. \ref{motiv:fig:3}, demonstrates that individual satellite links exhibit sharp peaks and rapid fluctuations due to satellite motion. At any given time, only a few satellites provide high quality links, and these favorable conditions are short lived. To capture broader trends, Fig. \ref{motiv:fig:4} presents the maximum SNR over time across multiple cities (Mumbai, Arizona, Sydney, and Quebec). The figure highlights significant spatial and temporal variability in achievable link quality, driven by orbital geometry and regional satellite density. This heterogeneity implies that a globally optimal handover strategy must account for both time evolution and geographic diversity, rather than relying on localized heuristics.

In addition, the availability of feasible satellites above a quality threshold varies across regions. As shown in Fig. \ref{motiv:fig:5}, the number of satellites satisfying a minimum SNR threshold differs significantly between cities, with regions like Arizona and Sydney typically having more options than Mumbai. This variation directly impacts the flexibility of handover decisions and the likelihood of redundant switching. Finally, the cumulative effect of these dynamics is reflected in the total number of switching instances per city, shown in Fig. \ref{motiv:fig:6}. A substantial number of handovers are observed across all regions, confirming that frequent switching is an inherent characteristic of LEO networks under conventional strategies. Therefore, efficient satellite handover requires a predictive, and optimization-based framework that balances link quality and handover cost.

\section{System Model and Problem Formulation}
\label{SC:system:model}
\begin{figure}[ht]
\centering
\includegraphics[scale=0.37]{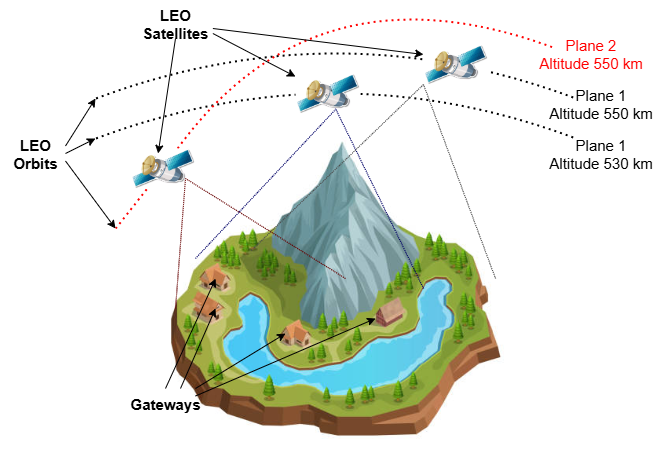}
\caption{The figure depicts a system model in which multiple LEO satellites orbit the Earth across various planes and altitudes, providing service to gateways distributed over different geographic regions on the Earth's surface.}
\label{fig:system model}
\end{figure}

Consider a region on the surface of the earth, in which there is a set of users $\mathcal{M}$; let $|\mathcal{M}| = M$. Each user is the set of devices in a home, office, community, etc., communicating with satellites via a single cluster-head, ground station, or base station (BS).
The users are served by the satellites in a LEO satellite constellation. Let the set of satellites in the constellation be $\mathcal{N}$ and $|\mathcal{N}| = N$.
Recall that for a given point on the earth's surface, the set of satellites that are visible is a periodic function, with the period being equal to the amount of time a satellite takes to orbit the earth. A LEO satellite typically takes between $90$ minutes and $2$ hours to complete an orbit around the earth. Consider a time horizon $\mathcal{T}=\{0,1, \ldots, T-1\}$, where $T \geq 2$ is an integer. Fig. \ref{fig:system model} illustrates the system model.

Let the demand for service from user $j$ in time slot $t$ be $D_t^j$. Since each user is the aggregate of a large number of devices, we assume that using the law of large numbers, the values of $D_t^j$ for each $t \in \mathcal{T}$ and $j \in \mathcal{M}$ can be approximately predicted in advance.
Let the data rate of user $j$ from satellite $i$ in time slot $t$ be $R_t^{i,j}$. Since the motion of the satellites is predictable, we assume that the values of $R_t^{i,j}$, $i \in \mathcal{N}$, $j \in \mathcal{M}$, $t \in \mathcal{T}$ are known. 

Let:
\begin{equation*}
\resizebox{252pt}{!}{$
X_t^{i,j} = \left\{ \begin{array}{ll}
    1, & \mbox{if user } j \mbox{ is associated with satellite } i \mbox{ in slot } t,  \\
    0, & \mbox{else.}
\end{array} \right.$}
\end{equation*}
We have the following constraints:
\begin{equation}
\label{EQ:binary:constraints}
X_t^{i,j} \in \{0,1\}, \, \forall i \in \mathcal{N}, j \in \mathcal{M}, t \in \mathcal{T}.    
\end{equation}

Let $H^{i,i^{\prime}}$\footnotemark{} be the cost of handing over a user from satellite $i$ to satellite $i^{\prime}$.
\footnotetext{It is a unitless relative penalty modeled to be directly proportional to the associated handover latency. Thus, larger values indicate higher handover latency and penalty.}
Note that this cost will be high (respectively, low) if $i$ and $i^{\prime}$ belong to different (respectively, the same) orbit(s). Also, no cost is incurred when a handover is not performed; hence, $H^{i,i} = 0$, $\forall i \in \mathcal{N}$. In each time slot $t$, each user $j$ must be associated with exactly one satellite, resulting in the following constraint:
\begin{equation}
\label{EQ:constraint:each:user:one:satellite}
\sum_{i \in \mathcal{N}} X_t^{i,j} = 1, \, \forall j \in \mathcal{M}, \forall t \in \mathcal{T}.
\end{equation}

Let the capacity (e.g., available bandwidth) of satellite $i \in \mathcal{N}$ be $C^i$. The following constraint captures the fact that the total amount of service provided by a satellite is at most its capacity:
\begin{equation}
\label{EQ:satellite:capacity:constraint}
\sum_{j \in \mathcal{M}} \frac{D_t^j }{R_t^{i,j}} X_t^{i,j} \leq C^i, \, \forall i \in \mathcal{N}, \forall t \in \mathcal{T}.    
\end{equation}
The total handover cost incurred in the $T$ time slots is:
\begin{equation}
\label{EQ:handover:cost}
\sum_{t \in \mathcal{T}} \sum_{i \in \mathcal{N}} \sum_{i^{\prime} \in \mathcal{N}} \sum_{j \in \mathcal{M}} H^{i,i^{\prime}} X_t^{i^{\prime},j} X_{(t-1)}^{i,j}.    
\end{equation}
The optimization horizon consists of the $T$ slots in $\mathcal{T}$, with handover costs incurred between consecutive slots. 
Our objective is to minimize the total handover cost in \eqref{EQ:handover:cost} subject to the constraints \eqref{EQ:binary:constraints}, \eqref{EQ:constraint:each:user:one:satellite}, and \eqref{EQ:satellite:capacity:constraint}.  

\begin{theorem}
    The satellite handover optimization problem described in \eqref{EQ:handover:cost} is NP-hard.
\end{theorem}

\begin{IEEEproof}
    We prove the NP-hardness of \eqref{EQ:handover:cost} by a polynomial-time reduction from the generalized assignment problem (GAP), which is known to be NP-hard \cite{fisher1986multiplier, cattrysse1992survey}.

    The GAP is defined as follows: Given a set of jobs $\mathcal{J} = \{1, \ldots, n\}$ and a set of agents $\mathcal{I} = \{1, \ldots, m\}$, each job $j \in \mathcal{J}$ must be assigned to exactly one agent $i \in \mathcal{I}$. Assigning job $j$ to agent $i$ incurs a cost $c_{ij}$ and consumes $a_{ij}$ units of the agent's capacity. Each agent $i$ has a capacity limit $b_i$. The objective is to minimize the total assignment cost subject to capacity constraints:
    \begin{align} \label{GAP_objective_function}
        \min \; &\sum_{i \in \mathcal{I}} \sum_{j \in \mathcal{J}} c_{ij} x_{ij}, \notag \\
        \text{s.t.} \; & \sum_{i \in \mathcal{I}} x_{ij} = 1, \;\;\;\;\;\;\;\;\; \forall j \in \mathcal{J}, \notag \\
        &\sum_{j \in \mathcal{J}} a_{ij} x_{ij} \leq b_i, \;\;\;\; \forall i \in \mathcal{I}, \notag \\
        & x_{ij} \in \{0,1\}, \;\;\;\;\;\;\;\;\; \forall i \in \mathcal{I}, j \in \mathcal{J},
    \end{align}
    where $x_{ij}$ is the assignment variable, and is equal to $1$ if agent $i$ performs job $j$, and $0$ otherwise. 

    We reduce an arbitrary instance of GAP to a special case of the satellite handover optimization problem. Construct a restricted instance of the satellite handover problem with two time slots, $T=2$, in which the serving satellite $i_0^j$ of each user $j$ at $t=0$ is fixed and given as part of the instance.
    
    At $t=1$, let the feasible satellites correspond to the GAP agents, i.e., $\mathcal{S}_1^j = \mathcal{I}, \forall j \in \mathcal{J}$. For each user--satellite pair, choose the demand and rate such that
    \[
    \frac{D_1^j}{R_1^{i,j}}=a_{ij},
    \]
    and set $C^i=b_i$. Hence, the satellite-capacity constraint becomes
    \[
    \sum_{j \in \mathcal{J}} a_{ij} x_{ij} \le b_i, \quad \forall i \in \mathcal{I},
    \]
    which is identical to the GAP resource constraint.
    
    Finally, define the handover cost from the fixed initial satellite $i_0^j$ to satellite $i$ as $H^{i_0^j,i} = c_{ij}$. Since $T=2$, each user makes one transition from $i_0^j$ to its selected satellite at $t=1$. Therefore, the total handover cost is
    \[
    \sum_{j \in \mathcal{J}} \sum_{i \in \mathcal{I}} H^{i_0^j,i} x_{ij} = \sum_{j \in \mathcal{J}} \sum_{i \in \mathcal{I}} c_{ij} x_{ij},
    \]
    which is exactly the GAP objective. The one-satellite-per-user constraint gives
    \[
    \sum_{i \in \mathcal{I}} x_{ij} = 1,\quad \forall j \in \mathcal{J}.
    \]
    
    Thus, every feasible solution of the constructed satellite problem corresponds to a feasible GAP solution with the same objective value, and vice versa. The construction is polynomial in the size of the GAP instance. Therefore, the satellite handover-cost minimization problem is NP-hard.
\end{IEEEproof}

\section{Branch-and-Price Framework} \label{methodology}

One of the standard approaches to efficiently solve the satellite handover optimization problem is the B\&P framework, which integrates CG to handle large-scale formulations. In this section, we present the baseline B\&P framework used to solve the above combinatorial problem.

\subsection{Reformulation of Problem} \label{reformulation problem}
The objective function of the satellite handover problem is inherently nonlinear due to the coupling between satellite assignments across consecutive time slots, where the handover cost depends on the joint selection of satellites at times $t$ and $t-1$. Such multiplicative decision terms make the problem computationally challenging to solve directly at scale. To enable efficient optimization, this nonlinear objective is reformulated into an equivalent linear representation by introducing structured decision variables that preserve the original cost semantics. A conventional time-indexed formulation treats each time slot independently and does not explicitly capture service continuity over time. Here, service continuity refers to the maintenance of a stable and uninterrupted communication session over time, with handovers occurring only as discrete and feasible transitions between satellites. This is particularly important because each user can be connected to only one satellite at any given time. To explicitly capture temporal dependencies and handover transitions, the problem is reformulated into a trajectory-based (route-based) representation \cite{liang2024optimizing}, where each decision corresponds to a complete sequence of satellite assignments over the horizon. 

In this reformulation, instead of defining binary variables that indicate which satellite serves a user at each time slot, a trajectory for a user is defined as an ordered sequence of satellites over the entire time horizon. A trajectory is locally feasible if it assigns exactly one visible, rate-feasible satellite to the user at each time slot. Global satellite-capacity constraints are enforced collectively in the master problem. Each feasible trajectory implicitly captures all handover events and their associated costs, which are accumulated along the sequence. The optimization problem then reduces to selecting exactly one feasible trajectory per user, subject to satellite capacity constraints. This reformulation yields a path-selection problem with exponentially many trajectories, motivating B\&P for efficient exploration.

For each user $j \in \mathcal{M}$, let a feasible handover trajectory be defined as, $p^j = \left(  i_0^j, i_1^j, \ldots, i_{T-1}^j\right)$, where $i_t^j \in \mathcal{N}$ represents the satellite serving user $j$ at time slot $t$. The feasibility of a trajectory is determined by local feasibility conditions associated with an individual user trajectory, such as visibility and rate requirements, while global capacity constraints are handled separately in the master problem. Each trajectory implicitly encodes all handover events across time, and its associated cost is given by $c_{p^j} = \sum_{t=1}^{T-1} H^{i_{t-1}^j, i_t^j}$. Let  $\mathcal{P}_j$ denote the set of all feasible trajectories for user $j$, which grows exponentially with the time horizon $T$ (on the order of $N^T$). 

\subsection{Master Problem}
The master problem minimizes the total handover cost while selecting exactly one trajectory per user and enforcing satellite capacity constraints at each time slot. In the trajectory-based formulation, each trajectory specifies the satellite assigned to a user over the entire time horizon, thereby providing a compact representation of the temporal coupling and enabling column generation over the exponentially large trajectory space.

Let $\lambda_{p^j}$ be a binary decision variable for each user $j$ and trajectory $p^j \in \mathcal{P}_j$, defined as:
\begin{equation*}
    \lambda_{p^j}=\begin{cases}
        1, & \text{if trajectory $p^j$ is selected for user $j$},\\ 
        0, & \text{otherwise}.
    \end{cases}
\end{equation*}
The assignment constraint ensures that exactly one trajectory is selected for each user, i.e., $\sum_{p^j \in \mathcal{P}_j} \lambda_{p^j} = 1, \; \forall j \in \mathcal{M}$. To link trajectory selection with time-indexed satellite usage, define the parameter:
\begin{equation*}
\resizebox{252pt}{!}{$
    a^{i}_{p^j,t}=\begin{cases}
        1, & \text{if trajectory $p^j$ assigns satellite $i$ to user $j$ at time $t$},\\ 
        0, & \text{otherwise}.
    \end{cases}$}
\end{equation*}
Using this parameter, the satellite capacity constraints at each time slot are given by:
\begin{equation} \label{new_capacity_constraint}
    \sum_{j \in \mathcal{M}} \sum_{p^j \in \mathcal{P}_j} 
    a^{i}_{p^j,t} \frac{D^j_t}{R^{i,j}_t} \lambda_{p^j} 
    \leq C^i, \quad \forall i \in \mathcal{N}, \; t \in \mathcal{T}.
\end{equation}
The resulting integer master problem is formulated as:
\begin{align} \label{integer_master_problem}
    \min \quad & \sum_{j \in \mathcal{M}} \sum_{p^j \in \mathcal{P}_j} c_{p^j} \lambda_{p^j}, \notag \\
    \text{s.t.} \quad 
    & \sum_{p^j \in \mathcal{P}_j} \lambda_{p^j} = 1, \quad \forall j \in \mathcal{M}, \notag \\
    & \sum_{j \in \mathcal{M}} \sum_{p^j \in \mathcal{P}_j} 
    a^{i}_{p^j,t} \frac{D^j_t}{R^{i,j}_t} \lambda_{p^j} 
    \leq C^i, \quad \forall i \in \mathcal{N}, \; t \in \mathcal{T}, \notag \\
    & \lambda_{p^j} \in \{0,1\}, \quad \forall p^j \in \mathcal{P}_j,\; j \in \mathcal{M}.
\end{align}

This formulation captures all time-coupled handover decisions within trajectory variables, ensuring consistency in handover transitions. However, the number of feasible trajectories grows exponentially with the time horizon, highlighting the combinatorial complexity of the problem. The linear relaxation of \eqref{integer_master_problem} is solved within the CG framework.

\subsection{Restricted Master Problem and LP Relaxation} \label{RMP_classical_BP}
Since the set of feasible handover trajectories $\mathcal{P}_j$ for each user $j \in \mathcal{M}$ is exponentially large, directly solving the master problem is computationally intractable. Therefore, an RMP is considered, in which only a subset of feasible trajectories $\mathcal{P}'_j \subseteq \mathcal{P}_j$ is included. The RMP is obtained by restricting the master problem to the currently generated trajectories and relaxing the integrality constraints, allowing $\lambda_{p^j} \in [0,1]$. The resulting formulation is:
\begin{align} 
    \min \quad & \sum_{j \in \mathcal{M}} \sum_{p^j \in \mathcal{P}'_j} c_{p^j} \lambda_{p^j}, \label{relaxed_restricted_master_problem} \\
    \text{s.t.} \quad 
    & \sum_{p^j \in \mathcal{P}'_j} \lambda_{p^j} = 1, \quad \forall j \in \mathcal{M}, \label{rmp_constraint_assignment} \\ 
    & \sum_{j \in \mathcal{M}} \sum_{p^j \in \mathcal{P}'_j} 
    a^{i}_{p^j,t} \frac{D^j_t}{R^{i,j}_t} \lambda_{p^j} 
    \leq C^i, \quad \forall i \in \mathcal{N}, \; t \in \mathcal{T}, \label{rmp_constraint_capacity} \\
    & \lambda_{p^j} \in [0,1], \quad \forall p^j \in \mathcal{P}'_j,\; j \in \mathcal{M}. \label{rmp_constraint_integrality}
\end{align}
The RMP \eqref{relaxed_restricted_master_problem} is a linear program, as each trajectory variable $\lambda_{p^j}$ represents a complete handover sequence with a precomputed cost $c_{p^j}$. Consequently, all temporal coupling is captured within individual trajectory costs, and both the objective function and constraints are linear in the decision variables. This structure enables efficient solution using standard LP solvers within the proposed framework.

At the root node, the restricted trajectory set $\mathcal{P}_j'$ for each user $j$ is initialized with a small pool of diverse feasible trajectories. In this classical B\&P framework, up to four seed trajectories are generated per user via a set of greedy heuristics: a rate-based greedy trajectory, where the user is assigned at every time slot to the visible satellite offering the highest instantaneous data rate; a second-best rate greedy trajectory, using the second-highest rate visible satellite at each slot; a minimum-handover greedy trajectory, which at each slot selects the visible satellite minimizing the handover cost from the previous slot's assignment; and, if fewer than four unique trajectories have been generated, additional randomly sampled feasible trajectories. Duplicate trajectories are discarded, so $\mathcal{P}_j'$ may contain fewer than four columns for some users. Thus, initially, $\mathcal{P}_j'$ contains a small set of heuristically diverse feasible trajectories for each user, ensuring feasibility of the RMP without incorporating any learning-based enhancements.

\subsection{Dual of the LP-Relaxed RMP}
Solving the LP relaxation of the RMP \eqref{relaxed_restricted_master_problem} yields a set of dual variables that play a central role in trajectory generation. Let $\pi_j \in \mathbb{R}$ be the dual variable associated with the assignment constraint \eqref{rmp_constraint_assignment} for user $j$, and let $\mu_t^i \geq 0$ be the dual variable associated with the capacity constraint \eqref{rmp_constraint_capacity} for satellite $i$ at time slot $t$. The dual of the LP-relaxed RMP is given by:
\begin{align} 
    \max_{\pi_j, \mu_t^i} \quad & \sum_{j \in \mathcal{M}} \pi_j + \sum_{i \in \mathcal{N}} \sum_{t \in \mathcal{T}} C^i \mu^i_t, \label{dual_LP} \\
    \text{s.t.} \quad 
    & \pi_j + \sum_{i \in \mathcal{N}} \sum_{t \in \mathcal{T}} 
    a^{i}_{p^j,t} \frac{D^j_t}{R^{i,j}_t} \mu^i_t 
    \leq c_{p^j}, \quad \forall j \in \mathcal{M}, \; p^j \in \mathcal{P}'_j, \label{dual_capacity_constraint} \\
    & \mu_t^i \geq 0, \quad \forall i \in \mathcal{N}, \; t \in \mathcal{T} \label{dual_variable_constraint}.
\end{align}
The dual formulation follows directly from standard LP duality applied to the relaxed RMP. Each assignment constraint gives rise to a free dual variable $\pi_j$, while each capacity constraint induces a nonnegative dual variable $\mu_t^i$. Each trajectory variable in the primal corresponds to a dual feasibility constraint, ensuring that the weighted contribution of selecting that trajectory does not exceed its cost.

The dual variable $\pi_j$ represents the marginal cost of assigning a trajectory to user $j$, and serves as a baseline for evaluating candidate trajectories. It can also be interpreted as a shadow price reflecting the minimum handover cost required to assign exactly one feasible trajectory to that user. The dual variable $\mu_t^i$ captures the marginal value of satellite capacity at time $t$, reflecting congestion in satellite resources. Higher values of $\mu_t^i$ penalize trajectories that utilize satellite $i$ at time $t$, thereby discouraging capacity-intensive assignments. The interpretation of dual variables as shadow prices for user assignment and resource capacity follows standard results from LP duality and CG theory \cite{barnhart1998branch, desrosiers2024branch, desrosiers2005primer}. These dual variables play a crucial role in the pricing problem, where they are used to compute the reduced cost of candidate trajectories and guide the generation of improving columns.

\subsection{Pricing Problem} \label{pricing_subsection_BP}
Using the dual variables obtained from \eqref{dual_LP}, the objective of the pricing problem is to identify, for each user $j \in \mathcal{M}$, a trajectory $p^j \in \mathcal{P}_j$ with negative reduced cost. The reduced cost of a trajectory $p^j$ is given by:
\begin{equation} \label{reduced_cost}
    \bar{c}_{p^j} = c_{p^j} - \pi_j - \sum_{i \in \mathcal{N}} \sum_{t \in \mathcal{T}} 
    a^{i}_{p^j,t} \frac{D^j_t}{R^{i,j}_t} \mu^i_t,
\end{equation}
which measures the net contribution of adding trajectory $p^j$ to the current RMP solution. A trajectory satisfies the corresponding dual constraint if its reduced cost is nonnegative. Conversely, a trajectory with negative reduced cost violates the dual constraint, implying that the current RMP solution is suboptimal with respect to the full trajectory space. Such trajectories are therefore candidates for inclusion in the RMP. So, a trajectory with $\bar{c}_{p^j} < 0$ corresponds to an improving column that can be added to the RMP. Therefore, the pricing problem for each user $j$ reduces to solving:
\begin{equation} \label{pricing_problem}
    \min_{p^j \in \mathcal{P}_j} \; \bar{c}_{p^j}.
\end{equation}
Substituting an expression of the trajectory cost $c_{p^j}$ defined in Section \ref{reformulation problem}, into \eqref{reduced_cost} and expanding over time, the reduced cost can be expressed as:
\begin{equation} \label{time_expanded_rc}
    \bar{c}_{p^j} = \sum_{t \in \mathcal{T}} H^{i_{t-1}^j, i_t^j}
    - \pi_j 
    - \sum_{t \in \mathcal{T}} \mu^i_t \frac{D^j_t}{R^{i,j}_t}.
\end{equation}

This formulation reveals that the pricing problem can be interpreted as a shortest path problem over a time-expanded graph, where nodes represent feasible satellite assignments at each time step, and edges capture handover costs. The node costs are adjusted using the dual variables $\mu_t^i$, which penalize the use of congested satellites. Due to the temporal coupling introduced by handover costs, the pricing problem can be efficiently solved using DP. At each time step, the DP recursively computes the minimum reduced cost of partial trajectories by combining transition costs and dual-adjusted node costs, while ensuring feasibility with respect to visibility and other constraints.

If a trajectory with negative reduced cost is identified, it is added to the RMP. Otherwise, no improving column exists, and the current LP solution is optimal for the RMP.

The pricing problem \eqref{pricing_problem} is a shortest-path problem on a
time-expanded graph. Define stage costs:
\begin{equation} \label{eq:stage}
  \eta_t^i(j,\mu) = -\mu_t^i \frac{D^j_t}{R_t^{i,j}},  \quad i \in \mathcal{V}_t^j,
\end{equation}
with $\eta_t^i = +\infty$ if $R_t^{i,j}= 0$. Here, $\mathcal{V}_t^j$ is the set of visible satellites for user $j$ at time $t$. The DP recursion is:
\begin{equation} \label{eq:DP}
  V_j(t, i) = \eta_t^i(j,\mu) + \min_{i' \in \mathcal{V}_{t-1}^j, V_j (t-1, i') <\infty}
    \bigl[V_j (t-1, i') + H^{i,i'} \bigr],
\end{equation}
initialized as $V_j(0, i) = \eta_0^i(j,\mu)$ for $i \in \mathcal{V}_0^j$. Here, $V_j(t, i)$ is the minimum accumulated pricing cost for user $j$ up to time slot $t$, assuming that the user is assigned to satellite $i$ at time $t$. After $T$ stages, the minimum reduced cost trajectory is recovered by backtracking through the argmin of $V_j (T-1, i) - \pi_j$ over all $i \in \mathcal{V}_{T-1}^j$. The complexity per user is $O(|\mathcal{V}|^2 T)$ where $|\mathcal{V}| \ll N$ is the mean visibility set size.

\subsection{Column Generation Procedure}

Following the pricing phase, all trajectories with negative reduced cost are added to the corresponding restricted trajectory sets $\mathcal{P}'_j$, and the RMP is re-solved in its LP-relaxed form. This iterative process of solving the RMP and augmenting it with improving trajectories continues until no feasible trajectory with negative reduced cost can be identified for any user. At this point, the CG procedure is said to have converged.

At convergence, all dual feasibility conditions over the full trajectory space are satisfied, implying that the current solution of the LP-relaxed RMP is optimal for the full (unrestricted) master problem. Consequently, the optimal objective value of the LP-relaxed RMP provides a valid lower bound (LB) on the optimal integer handover cost.

By strong duality of linear programming \cite{bertsimas1997introduction}, the optimal values of the primal RMP and its dual coincide, and the absence of negative reduced cost columns provides the standard CG optimality certificate. Formally, let $z_P^*$ denote the optimal objective value of the LP-relaxed RMP, and $z_D^*$ denote the optimal value of its dual. By strong duality of LP, where the minimization is performed subject to the primal feasibility constraints \eqref{rmp_constraint_assignment}, \eqref{rmp_constraint_capacity}, \eqref{rmp_constraint_integrality} and the maximization is performed subject to the dual feasibility constraints in \eqref{dual_capacity_constraint}, and \eqref{dual_variable_constraint},
\begin{align*}
    z_P^* 
    &= \min_{\lambda_{p^j} \in [0,1]} \sum_{j \in \mathcal{M}} \sum_{p^j \in \mathcal{P}'_j} c_{p^j} \lambda_{p^j} \\
    &= \max_{\pi_j, \mu_t^i} \left( \sum_{j \in \mathcal{M}} \pi_j + \sum_{i \in \mathcal{N}} \sum_{t \in \mathcal{T}} C^i \mu_t^i \right) = z_D^*.
\end{align*}

This primal–dual equality justifies the reduced cost optimality condition used in the pricing problem and certifies convergence of the CG procedure. After convergence, the algorithm proceeds to integrality verification and branching within the B\&P framework.

\subsection{Branch and Bound Integration}
After convergence of the CG procedure at a given B\&B node, the LP solution of the RMP is examined for integrality. If the optimal solution satisfies $\lambda_{p^j} \in \{0, 1\}$ for all $p^j \in \mathcal{P}'_j$ and all $j \in \mathcal{M}$, it corresponds to a feasible integer solution that selects exactly one trajectory per user, and its objective value provides a valid UB on the optimal handover cost.

If the LP solution remains fractional, a branching step is performed to enforce integrality. Branching is carried out on a fractional assignment decision, namely whether a user $j$ is assigned to a satellite $i$ at time slot $t$. Let $x_{t,i}^j = \sum_{p^j \in \mathcal{P}'_j} a_{p^j, t}^i \lambda_{p^j}$ be the implied assignment of user $j$ to satellite $i$ at time $t$ in the LP solution. If $x_{t,i}^j \in (0,1)$, two child nodes are created: one enforcing $x_{t,i}^j = 1$, which restricts admissible trajectories to those assigning satellite $i$ to user $j$ at time $t$, and the other enforcing $x_{t,i}^j = 0$, which excludes all such trajectories. At each B\&B node, the branching restrictions are incorporated into the pricing subproblem by restricting the admissible satellite assignments at the corresponding user-time coordinates. Thus, all newly generated trajectories satisfy the branching constraints. The RMP is then re-optimized at each node using CG.

During the B\&B process, pruning is employed to eliminate nodes that cannot yield an improved integer solution. At each node, the optimal objective value of the LP-relaxed RMP provides a valid LB on the achievable objective value within that node, while the best integer-feasible solution found so far provides a global UB. If $LB \geq UB$, the node is pruned, as no solution in its feasible region can improve upon the current incumbent. The algorithm terminates when all nodes in the B\&B tree have either been explored or pruned.

\begin{algorithm}[t]
\caption{Branch-and-Price Framework}
\label{alg:bp_cg}
\begin{algorithmic}[1]
\renewcommand{\algorithmicrequire}{\textbf{Input:}}
\renewcommand{\algorithmicensure}{\textbf{Output:}}

\REQUIRE 
$\mathcal{M}, \mathcal{N}, \mathcal{T}, H^{i,i'}, D_t^j, R_t^{i,j}, C^i$

\ENSURE 
Optimal integer solution $\lambda^*$ and objective value $UB$

\STATE Generate an initial set of feasible trajectories $\mathcal{P}'_j$ for each user $j \in \mathcal{M}$ using a heuristic
\STATE Initialize root node with $\mathcal{P}'_j$
\STATE Set global upper bound $UB \leftarrow +\infty$
\STATE Initialize node list $\mathcal{L}$ with root node

\WHILE{$\mathcal{L}$ is not empty}

    \STATE Select, process and remove a node from $\mathcal{L}$

    \REPEAT
        \STATE Solve LP-relaxed RMP and obtain $(\lambda^*, \pi_j, \mu_t^i)$
        \STATE Solve pricing problem \eqref{pricing_problem} for each user $j$:
        \STATE Let $\bar{c}_{p^j}$ denote the reduced cost of trajectory $p^j$
        
        \IF{$\exists \, p^j$ such that $\bar{c}_{p^j} < 0$}
            \STATE Add $p^j$ to $\mathcal{P}'_j$
        \ENDIF
        
    \UNTIL{no trajectory with negative reduced cost is found}

    \STATE Compute lower bound $LB \leftarrow$ objective value of RMP

    \IF{$LB \geq UB$}
        \STATE Prune node; continue
    \ENDIF

    \IF{$\lambda^*$ is integral}
        \STATE Update $UB \leftarrow LB$; Store the solution
    \ELSE
        \STATE Create two child nodes:
        \STATE \quad Node 1: enforce $x_{t,i}^j = 1$
        \STATE \quad Node 2: enforce $x_{t,i}^j = 0$
        
        \STATE Add child nodes to $\mathcal{L}$
    \ENDIF

\ENDWHILE

\STATE \textbf{return} best integer solution and $UB$

\end{algorithmic}
\end{algorithm}


\section{RL-Based Multiple-Column Selection - Branch-and-Price (MCS-B\&P) } \label{true_RL}

The optimization problem \eqref{integer_master_problem} is exactly solvable via B\&P, but the pricing subproblem of classical B\&P adds only a single column per user per CG iteration, which is often the computational bottleneck for the constellation sizes considered here. We propose \emph{MCS-B\&P}, an exact B\&P variant that accelerates CG by selecting multiple promising columns per iteration via an RL policy, following the multiple-column selection strategy of \cite{yuan2024reinforcement} for generic integer programs. The work \cite{yuan2024reinforcement} evaluates its RL-based selection policy on cutting-stock and graph-coloring benchmarks, reporting faster LP convergence than single-column CG; we specialize the same idea to the per-user handover trajectory structure of the optimization problem \eqref{integer_master_problem}. A forced-optimal column rule ensures that the accelerated procedure never sacrifices the exactness of classical B\&P.

At the root node, $\mathcal{P}'_j$ is initialized identically for classical B\&P and MCS-B\&P using the same routine. The algorithms differ only in the columns added to $\mathcal{P}'_j$ during CG.

\subsection{Candidate Column Generation} \label{ssec:candidate_gen}

At each CG iteration, candidate generation proceeds in two phases:

\emph{Phase 1 (all $M$ users):} For every user $j$, one exact DP pass over \eqref{eq:DP} (perturbation index $0$) returns the true minimizer $p^j_{(1)}=p^{j\star}$ of \eqref{pricing_problem} — identical to the exact pricing step used in classical B\&P.

\emph{Phase 2 (priority queue):} Only the $q$ users with the largest $|\bar{c}_{p^j_{(1)}}|$ (i.e.\ those furthest from local optimality, and hence with the most room for the RMP to improve) receive $n_{\mathrm{runs}}-1$ additional diversified DP passes. Each diversified pass re-solves \eqref{eq:DP} under a perturbed edge cost that penalizes the satellite selected at the corresponding time slot by the initial unperturbed minimum reduced cost trajectory,
\begin{equation}\label{eq:perturb}
  \bar{H}^{i,i'} = H^{i,i'} + \delta_f \cdot \mathbf{1}\!\left[i = i^{(f-1)}_{t}\right],
\end{equation}
with an escalating penalty scale $\delta_f$ across passes $f=1, \ldots,n_{\mathrm{runs}}-1$, nudging successive passes toward structurally different — often cross-shell or cross-plane — trajectories. Passes yielding a non-negative reduced cost or a duplicate assignment are discarded. The per-user candidate set is $\mathcal{F}_j =\{p^j_{(1)}, \ldots, p^j_{(n_j)}\}$, $n_j \le n_{\mathrm{runs}}$, and the pooled candidate set offered to the RL policy is $\mathcal{F} = \bigcup_j \mathcal{F}_j$, $n_{\mathrm{cand}} = |\mathcal{F}|$.

\begin{proposition} \label{prop:cand_cost}
At most $M+q(n_{\mathrm{runs}}-1)$ DP pricing passes are performed
during candidate generation at each CG iteration.
\end{proposition}
\begin{IEEEproof}
Phase 1 performs one pass for each user ($M$ total), while Phase 2 performs at most $n_{\mathrm{runs}}-1$ additional passes for each of the $q$ prioritized users. Summing gives $M + q(n_{\mathrm{runs}}-1)$; no other DP solves occur in candidate generation.
\end{IEEEproof}


\subsection{Markov Decision Process Formulation} \label{ssec:mdp}
We formulate the multiple-column selection at CG generation $\ell$ as a Markov decision process (MDP) following \cite{yuan2024reinforcement}.

\paragraph{State $s$} \label{para:state_MDP}
The state comprises a bipartite graph $G = (\mathcal{U} \cup \mathcal{K}, E)$ and a global feature vector. The constraint node set $\mathcal{U}$ contains two families:
\begin{equation} \label{eq:mcs_constraint_nodes}
    \mathcal{U} = \underbrace{\{u_j : j \in \mathcal{M}\}}_{\text{assignment nodes, dual } \pi_j} \ \cup \ \underbrace{\{u_{i,t} : |\mu^i_t| \ge \varepsilon_\mu \}}_{\text{capacity nodes, dual } \mu^i_t},
\end{equation}
where $\varepsilon_\mu$ is a small positive constant. The column node set $\mathcal{K} = \mathcal{K}_{\mathrm{exist}} \cup \mathcal{F}$ contains the trajectories currently in the RMP and the candidate pool $\mathcal{F}$. An edge $(u,v) \in E$ exists whenever column $v$ contributes to constraint $u$: an assignment edge $(u_j,v)$ if $v$ corresponds to user $j$, and a capacity edge $(u_{i,t},v)$ if trajectory $v$ uses satellite $i$ at time slot $t$.

Each column node has a $7$-dimensional feature vector comprising its reduced cost, normalized trajectory cost, mean load fraction, handover fraction, and fractions of intra-plane, intra-shell, and inter-shell handovers. Each assignment node is represented by the $4$-dimensional feature vector,
\[
y_{u_j} = \Big[\, \bar{D}^j / D_{\max},\ |\pi_j| / \pi_{\max},\
\overline{|\mathcal{V}^j_{\cdot}|} / \mathcal{V}_{\max},\ 0 \,\Big],
\]
where $\bar{D}^j = \frac{1}{T} \sum_{t=0}^{T-1} D_t^j$ is the average demand of user $j$, $D_{\max} = \max_{j \in \mathcal{M}, t \in \mathcal{T}} D^j_t$ is the maximum per-slot demand, $|\mathcal{V}^j_{\cdot}| = \frac{1}{T} \sum_{t=0}^{T-1} |\mathcal{V}^j_t|$ is the average number of visible satellites for user $j$, and $\mathcal{V}_{\max} = \max_{j \in \mathcal{M}, t \in \mathcal{T}} |\mathcal{V}^j_t|$ is the maximum per-slot visible-satellite count. Each active capacity node has the feature vector,
\[
y_{u_{i,t}} = \Big[\, U_{i,t},\ |\mu^i_t| / \mu_{\max},\ U_{i,t},\
\mathrm{shell}(i) \Big],
\]
where $U_{i,t} = \frac{1}{C^i}\sum_{k \in \mathcal{M}}\frac{D_t^k}{R_t^{i,k}}$ is the predicted utilization of satellite $i$ at slot $t$, and $\mathrm{shell}(i)$is a normalized orbital-shell index. The global feature vector is $\big[ M/M_{\max}, N/N_{\max}, \ell/\ell_{\max}, z_{\mathrm{LP}}/z_0, U_{samp} \big]$, 
where $M_{\max}=200$ and $N_{\max}=5000$ are fixed normalization constants representing the maximum user and satellite counts considered in the implementation, respectively. Here, $M$ and $N$ denote the numbers of users and satellites in the instance, $\ell_{\max}=300$ is the maximum number of CG iterations, $z_{\mathrm{LP}}$ is the current RMP objective value, $z_0$ is its value at CG iteration $0$, and $U_{\mathrm{samp}}$ is the maximum normalized aggregate satellite utilization across the sampled time slots.

\paragraph{Action $a$}
At each CG iteration, the agent selects a $k$-subset of the $n_{\mathrm{cand}}$ candidate columns in $\mathcal{F}$ without replacement. Thus, the action space contains $|\mathcal{A}| = \binom{n_{\mathrm{cand}}}{k}$ possible unordered subsets. This formulation follows the multiple-column selection setting in \cite{yuan2024reinforcement}.

\paragraph{Transition}
The selected $k$ columns are added to $\mathcal{P}'_j$, the RMP is re-solved, and dual variables update deterministically.

\paragraph{Reward $r$}
Let $z_{\ell-1},z_{\ell}$ be the RMP objective before and after adding the columns selected at CG iteration $\ell$, $z_0$ be the initial RMP objective, and $\alpha$ be the sensitivity control parameter. We define the normalized objective improvement as,
\begin{equation*} 
  \Delta z_{\ell}=\frac{z_{\ell-1}-z_{\ell}}{\max(|z_0|,1)}, \qquad
  \widehat{\Delta z}_{\ell}=\tanh(\alpha \Delta z_{\ell})\in(-1,1).
\end{equation*}
The diversity of the selected column set $\mathcal{F}_s$ is measured using the average Jaccard distance between the sets of occupied $(i,t)$ pairs of the selected trajectories:
\begin{equation*} 
    \mathrm{div}(\mathcal{F}_s) = \frac{1}{|\mathcal{F}_s|(|\mathcal{F}_s|-1)} \sum_{\substack{u,v \in \mathcal{F}_s \\ u \ne v}} \big(1 - J(u,v)\big),
\end{equation*}
where $J(u,v)$ is the Jaccard similarity between the sets of occupied $(i,t)$ pairs of trajectories $u$ and $v$. Thus, larger values of $\mathrm{div}(\mathcal{F}_s)$ correspond to more structurally diverse selected columns. The normalized dual movement is defined as,
\begin{equation*}
  \mathrm{dp}_{\ell} = \mathrm{clip} \! \left( \frac{\lVert \mu_{\ell} - \mu_{\ell-1} \rVert_1}{\lVert \mu_{\ell-1} \rVert_1 + \epsilon_{clip}}, 0, 1 \right),
\end{equation*}
where $\epsilon_{clip}$ is a small constant. The immediate reward for the MCS policy adapted here at CG iteration $\ell$ is then,
\begin{equation}\label{eq:reward}
  r_{\ell} = \widehat{\Delta z}_{\ell} + \beta \, \mathrm{div}(\mathcal{F}_s) + 0.1\,\mathrm{dp}_{\ell} - \big(1-\widehat{\Delta z}_{\ell} \big),
\end{equation}
with $\alpha=300$ and $\beta=0.02$. The final term penalizes iterations with limited RMP progress, while avoiding a fixed negative offset that would dominate the reward during the later stages of CG, when objective improvements naturally become small.

\subsection{Neural Network Architecture}
The MCS-B\&P policy is implemented as an actor--critic network consisting of a bipartite graph encoder, an actor decoder, and a critic decoder.

\subsubsection{Bipartite Graph Encoder}
The bipartite graph $G$ is encoded using $B_1$ alternating graph isomorphism network (GIN) layers \cite{xu2018powerful}. The initial column, constraint, and global features are first projected into a common hidden dimension. Let $l_i=0, \ldots, B_1-1$ denote the GIN layer index. At each layer, constraint information is first propagated to the column nodes, followed by propagation from the updated column nodes to the constraint nodes. For a column node $v\in\mathcal{K}$,
\begin{equation*}
  \mathbf{z}_v^{(l_i)} = \operatorname{MLP}_{\mathcal{U} \to \mathcal{K}}^{(l_i)} \!\left( \mathbf{h}_v^{(l_i)} + \sum_{u \in \mathcal{Q}(v)} \mathbf{h}_u^{(l_i)} \right),
\end{equation*}
\begin{equation*}
  \mathbf{h}_v^{(l_i+1)} = \operatorname{LeakyReLU} \! \left(\mathbf{z}_v^{(l_i)} \right) + \mathbf{h}_v^{(l_i)}.
\end{equation*}
The updated column embeddings are then propagated to constraint nodes: for $u\in\mathcal{U}$,
\begin{equation*}
  \mathbf{z}_u^{(l_i)} = \operatorname{MLP}_{\mathcal{K} \to \mathcal{U}}^{(l_i)} \! \left( \mathbf{h}_u^{(l_i)} + \sum_{v \in \mathcal{Q}(u)} \mathbf{h}_v^{(l_i+1)} \right),
\end{equation*}
\begin{equation*}
  \mathbf{h}_u^{(l_i+1)} = \operatorname{LeakyReLU} \! \left(\mathbf{z}_u^{(l_i)} \right) + \mathbf{h}_u^{(l_i)}.
\end{equation*}
Here $\mathcal{Q}(v) \subset \mathcal{U}$ and $\mathcal{Q}(u) \subset \mathcal{K}$ denote the neighboring constraint and column nodes, respectively, $\mathbf{z}_v^{(\cdot)} / \mathbf{z}_u^{(\cdot)}$ is the pre-activation MLP output, and $\mathbf{h}_v^{(\cdot)} / \mathbf{h}_u^{(\cdot)}$ is the node embedding. Each $\operatorname{MLP}^{(l_i)}_{\bullet \to \bullet}$ is a two-layer network with a ReLU hidden nonlinearity, and both GIN transformations use the standard self-weighting coefficient. Each transformation's output is passed through a LeakyReLU activation and combined with its layer-input via a residual connection. Thus every GIN layer performs the sequential information flow $\mathcal{U} \to \mathcal{K} \to \mathcal{U}$, letting column embeddings first incorporate their constraint neighbors' information, and constraint embeddings subsequently incorporate the just-updated column information.

\subsubsection{Candidate Interaction and Actor Decoder}
The candidate-column embeddings obtained from the bipartite encoder are further processed by a graph attention network (GAT) \cite{velivckovic2017graph} defined on a complete graph whose nodes are the candidate columns. Thus, each candidate can exchange information with all other candidates. The GAT produces an additional candidate embedding $\mathbf{h}_v^{\mathrm{GAT}}$, which is concatenated with the original bipartite embedding $\mathbf{h}_v^{(B_1)}$ and passed through an MLP to obtain a merged embedding:
\begin{equation*}
    \tilde{\mathbf{h}}_v = \operatorname{MLP}_{actor} \left( [\mathbf{h}_v^{(B_1)}; \mathbf{h}_v^{\mathrm{GAT}}] \right).
\end{equation*}
For each candidate $v$, the actor then produces an unnormalized selection logit from the merged embedding:
\begin{equation*}
    \zeta_v = \nu_{clip} \tanh \left( \mathbf{w}_o^\top \operatorname{ReLU}(W_o\tilde{\mathbf{h}}_v) \right),
\end{equation*}
where $\nu_{clip} = 10$ is the clipping coefficient, and $W_o$ and $\mathbf{w}_o$ are learnable parameters. The resulting logits $\{\zeta_v\}_{v \in \mathcal{F}}$ define the candidate-selection policy over the pooled candidate set $\mathcal{F}$. During training, $k$ candidates are sampled without replacement using Gumbel-Top-$k$ sampling \cite{kool2019stochastic}, which is distributionally equivalent to sequential sampling without replacement under the Plackett--Luce model \cite{plackett1975analysis}. During greedy evaluation, the $k$ candidates with the largest raw logits $\zeta_v$ are selected directly, without adding Gumbel noise.

\subsubsection{Critic Decoder}
The critic estimates the state value $\Gamma(s)$ by separately average-pooling the embeddings of existing RMP columns ($\bar{\mathbf{h}}_{\mathrm{exist}}$), candidate columns ($\bar{\mathbf{h}}_{\mathrm{cand}}$), and constraint nodes ($\bar{\mathbf{h}}_{\mathrm{cons}}$). These three pooled representations are concatenated with the global embedding $\mathbf{h}_g$ and passed through a three-layer MLP:

\begin{equation}\label{eq:critic}
    \Gamma(s) = \operatorname{MLP}_{\mathrm{critic}} \left( [ \bar{\mathbf{h}}_{\mathrm{exist}}; \bar{\mathbf{h}}_{\mathrm{cand}}; \bar{\mathbf{h}}_{\mathrm{cons}}; \mathbf{h}_{g} ]\right).
\end{equation}
If any of the three column/constraint sets is empty at a given CG iteration, its pooled representation is replaced by a zero vector of the same dimension as $\mathbf{h}_g$ rather than an undefined empty-set average, ensuring \eqref{eq:critic} remains well-defined at every CG iteration.

\subsection{Pre-training via PPO}
The policy is pre-trained directly on the target problem instance to adapt it to the constellation geometry, visibility structure, and demand distribution before deployment.

Pre-training consists of $30$ independent episodes, each initialized from a fresh deep copy of the shared initial trajectory set and run for $\Omega=12$ CG steps. To provide a richer candidate pool for policy learning, each candidate-generation phase allows at most four DP passes. The pre-training time is excluded from the B\&B solve-time budget, and the forced-optimal column rule described later is also applied during pre-training.

At the end of each episode, the collected transitions are used to update the policy and critic via PPO. The return for transition ($\ell$) is estimated using Monte Carlo returns, without generalized advantage estimation (GAE):
\begin{equation*}
    Y_{\ell} = \sum_{b=0}^{\ell_{\max}-\ell}\gamma^b r_{\ell+b}, \qquad \gamma=0.9,
\end{equation*}
where $\ell_{\max}$ denotes the final transition index of the episode. The advantage is computed as $A_{\ell} = Y_{\ell} - \Gamma_{\phi}(s_{\ell})$, where $\Gamma_{\phi}(s_{\ell})$ is the critic's estimated state value. For episodes containing multiple transitions, the advantages are standardized as,
\begin{equation*}
\hat A_{\ell} = \frac{A_{\ell} - \bar A}{\sigma_A + \epsilon_A}, 
\end{equation*}
where $\bar A $and $\sigma_A$ are the mean and standard deviation of the episode's advantages, respectively, and $\epsilon_A = 10^{-8}$ prevents division by zero.

Each episode is followed by one PPO update comprising $n_{\mathrm{epochs}}=4$ passes over its transitions. Each pass performs an individual gradient step for every valid transition, yielding at most $n_{\mathrm{epochs}} \Omega$ optimizer steps per episode. For a selected $k$-subset $\mathcal{F}_s$, its log-probability is evaluated under the Plackett--Luce model using a fixed ordering of the selected candidates:
\begin{equation*}
  \log P(\mathcal{F}_s) = \sum_{v \in \mathcal{F}_s} \Big[ \zeta_v - \log\!\sum_{u \in \mathcal{F}'(v)} \!\exp(\zeta_u)\Big],
\end{equation*}
where $\zeta_v$ is the logit of candidate $v$, and $\mathcal{F}'(v)$ contains the candidates remaining at the sequential selection step corresponding to $v$. Since the selected subset is unordered, evaluating only one of its $k!$ possible orderings yields a lower bound on its exact log-probability under the Plackett--Luce model.

The PPO probability ratio is
\begin{equation*}
    \rho_{\ell} = \exp! \left( \log P_{\theta} (\mathcal{F}_s) - \log P{\theta_{\mathrm{old}}}(\mathcal{F}_s) \right),
\end{equation*}
where $\theta{\mathrm{old}}$ denotes the policy parameters used to collect the episode. The joint policy--critic loss is
\begin{align}\label{eq:ppoloss2}
  \text{Loss}(\theta,\phi) =& -\mathbb{E}_{\ell} \Big[ \min \big( \rho_{\ell} \hat A_{\ell}, \mathrm{clip}(\rho_{\ell}, 1 {-} \epsilon, 1 {+} \epsilon) \hat A_{\ell} \big) \Big] \notag \\
  &+ c_1\,\mathbb{E}_{\ell} \big[(\Gamma_\phi(s_{\ell}) - Y_{\ell})^2 \big] - c_2\,\mathbb{E}_{\ell}[\mathcal{H}(\pi_\theta)],
\end{align}
where $\epsilon=0.2$, $c_1=0.5$, and $c_2=0.01$. The expectations are taken over the episode transitions, and $\mathcal{H}(\pi_{\theta})$ denotes the policy entropy, approximated by the softmax entropy over all candidate logits. The Adam optimizer uses a learning rate of $10^{-3}$, $\epsilon_{\mathrm{Adam}}=10^{-5}$, and gradient-norm clipping at $0.5$. After pre-training, the learned policy is kept fixed during the subsequent B\&B solve, with no further online policy updates.

\begin{table}[t]
  \caption{MCS-B\&P Hyperparameters}
  \label{tab:mcsbp_hyper2}
  \centering
  \begin{tabular}{lcl}
    \toprule
    Symbol & Value & Description \\
    \midrule
    $n_{\mathrm{cand}}$  & 20 & \texttt{MCS\_N\_CANDS} \\
    $k$                        & 10 & \texttt{MCS\_K\_SELECT} \\
    $q$                        & 10 ($=k$) & \texttt{MCS\_PRIORITY\_Q} \\
    $\alpha$                   & 300 & Reward $\tanh$ sensitivity \\
    $\beta$                    & 0.02 & Reward diversity weight \\
    $d_h$                      & 128 & Hidden dim.\ (encoder/decoder) \\
    $B_1$                      & 3   & GIN layers \\
    $\texttt{PRETRAIN\_ITERS}$                        & 30  & Pretraining episodes \\
    $\Omega$                        & 12  & CG steps per pretraining episode \\
    $\gamma,\epsilon$          & 0.9, 0.2 & PPO discount/clip \\
    \bottomrule
  \end{tabular}
\end{table}

\paragraph{Forced-Optimal-Column Rule}
At each CG iteration, MCS-B\&P adds to the RMP both the columns selected by the policy and the exact minimum reduced cost trajectory for every user. Let $p^j_{(1)}$ denote the minimum reduced cost trajectory obtained by solving the pricing problem for user $j$. The forced-column set is
\begin{equation}\label{eq:forced2}
    \mathrm{Forced}(\mathcal{F}) = \{p^j_{(1)} : j \in \mathcal{M}\}.
\end{equation}
Accordingly, the trajectory set for user $j$ is updated as $\mathcal{P}'^{\,(\ell+1)}_j = \mathcal{P}'^{\,(\ell)}_j \cup \{p^j_{(1)}\} \cup (\mathcal{F}_s \cap \mathcal{P}_j)$. Thus, the exact minimum reduced cost trajectory of every user is always included in the RMP, while the RL policy can contribute additional candidate columns.

\begin{algorithm}[t]
\caption{MCS-B\&P Column-Generation Loop}
\label{alg:mcsbp}
\begin{algorithmic}[1]
\renewcommand{\algorithmicrequire}{\textbf{Input:}}
\renewcommand{\algorithmicensure}{\textbf{Output:}}

\REQUIRE trained policy $\pi_\theta$, node's restricted trajectory sets $\{\mathcal{P}'_j\}_{j\in\mathcal{M}}$

\REPEAT

\STATE Solve the RMP to obtain $(\lambda^*,\{\pi_j\}_{j \in \mathcal{M}},\{\mu_t^i\})$

\STATE $\mathrm{NewColumn} \leftarrow \mathrm{False}$

\FOR{each user $j \in \mathcal M$}
    \STATE Generate candidate pool $\mathcal{F}_j$ (Section \ref{ssec:candidate_gen}) by solving DP pricing 
\ENDFOR

\STATE $\mathcal{F} \leftarrow \bigcup_{j \in \mathcal M} \mathcal{F}_j$ (Section \ref{ssec:candidate_gen})

\STATE Build state $s$ from $\mathcal{F}$ and the current RMP

\STATE Select $k$ candidates $\mathcal{F}_s \leftarrow \operatorname{TopK}_{v \in \mathcal{F}} \zeta_v$
using the trained policy

\STATE
$\mathrm{Forced}(\mathcal{F}) \leftarrow \{p^j_{(1)} : j \in \mathcal{M}\}$

\FOR{each user $j \in \mathcal{M}$}

    \FOR{each trajectory $p \in \mathcal{F}_s \cup \mathrm{Forced}(\mathcal{F})$ belonging to user $j$}
        \IF{$p\notin\mathcal P'_j$}
            \STATE Add $p$ to $\mathcal P'_j$
            \STATE $\mathrm{NewColumn}\leftarrow\mathrm{True}$
        \ENDIF
    \ENDFOR

\ENDFOR

\UNTIL{$\mathrm{NewColumn}=\mathrm{False}$}

\STATE Apply branching and pruning as in Algorithm~1

\end{algorithmic}
\end{algorithm}

\subsection{Complexity Analysis}\label{ssec:mcsbp_complexity}
\begin{proposition} \label{prop:mcsbp_time}
Each CG iteration of Algorithm \ref{alg:mcsbp} runs in
\begin{align}
    T_{\mathrm{LP}} &\big(n_{\mathrm{exist}}, \, M + N T \big) + O\!\left( \big[ M + q(n_{\mathrm{runs}} - 1 ) \big] T \mathcal{V}_{max}^2 \right) \notag \\
    &+ O\big( B_1(n_{\mathrm{exist}} + n_{\mathrm{cand}}) T \big) + O \big(n_{\mathrm{cand}}^2 \big)
\end{align}
where $T_{\mathrm{LP}}(n_{\mathrm{var}},n_{\mathrm{con}})$ denotes the solver-dependent cost of one RMP LP solve with $n_{\mathrm{var}}$ variables and $n_{\mathrm{con}}$ constraints, and $n_{\mathrm{exist}}$ denotes the number of existing columns included in the bipartite graph. The remaining three terms correspond, respectively, to DP candidate generation, the bipartite GIN encoder, and the candidate-interaction GAT. The RMP term $T_{\mathrm{LP}} (n_{\mathrm{exist}}, M + N T)$ is common to classical single-column B\&P when evaluated at the same RMP column set. The remaining terms therefore characterize the additional computational overhead introduced by MCS-B\&P.
\end{proposition}
\begin{IEEEproof}
At most $M + q(n_{\mathrm{runs}}-1)$ DP pricing solves are performed in each candidate-generation phase by Proposition \ref{prop:cand_cost}. Since each DP solve requires $O(T \mathcal{V}_{max}^2)$ time, candidate generation requires $O\!\left( \big[ M + q(n_{\mathrm{runs}} - 1 ) \big] T \mathcal{V}_{max}^2 \right)$.

The RMP contains $n_{\mathrm{exist}}$ variables, one for each trajectory currently included in the RMP. It contains $M$ assignment constraints and $n_{\mathrm{cap}}$ capacity constraints, where $n_{\mathrm{cap}}\le NT$ denotes the number of satellite-time capacity constraints represented in the RMP. Thus, in worst case, its solution cost is represented by $T_{\mathrm{LP}} (n_{\mathrm{exist}}, M + NT)$.

The bipartite graph contains $O((n_{\mathrm{exist}} + n_{\mathrm{cand}})T)$ edges because each trajectory has at most one assignment edge and at most one capacity edge per time slot. With fixed hidden dimensions and a fixed number $B_1$ of GIN layers, the encoder therefore requires $O\!\left( B_1(n_{\mathrm{exist}} + n_{\mathrm{cand}}) T \right)$ time.

Finally, the candidate-interaction graph is complete over the $n_{\mathrm{cand}}$ candidate nodes. Hence it contains $O(n_{\mathrm{cand}}^2)$ edges. With fixed hidden dimensions and a fixed number of GAT layers, the candidate-interaction component requires $O(n_{\mathrm{cand}}^2)$ time. Combining these terms gives the stated bound.
\end{IEEEproof}

\subsection{Convergence and Optimality Analysis} \label{ssec:mcsbp_conv}
\begin{lemma} \label{lem:monotone}
For column sets $\mathcal{P}'_j \subseteq \mathcal{P}''_j$ for all $j \in \mathcal{M}$, the optimal objective values of the corresponding LP-relaxed RMPs satisfy
\[
z_{\mathrm{LP}} (\{\mathcal{P}''_j\}) \le z_{\mathrm{LP}}(\{\mathcal{P}'_j\}).
\]
\end{lemma}
\begin{IEEEproof}
Every feasible solution of the RMP over $\{\mathcal{P}'_j\}$ remains feasible over $\{\mathcal{P}''_j\}$ by assigning zero to all newly added columns. Hence, the feasible region of the latter RMP contains that of the former. Since the RMP is a minimization problem, its optimal objective value cannot increase.
\end{IEEEproof}

\begin{theorem} \label{thm:convergence}
Assume that the CG procedure is allowed to run to completion without time-budget truncation and that the pricing problem is solved exactly. Then the MCS-B\&P CG loop terminates after finitely many iterations at an RMP whose LP objective equals the optimal value of the node's full LP relaxation, $z^\star_{\mathrm{LP}}$. Consequently, if the subsequent B\&B procedure is also allowed to run to completion, the resulting B\&P algorithm returns a global optimum of Problem \eqref{integer_master_problem}.
\end{theorem}

\begin{IEEEproof}
For each user $j$, the set $\mathcal P_j$ of feasible trajectories is finite. Hence, the sequence of restricted trajectory sets generated by the CG procedure can strictly increase only finitely many times.

At every CG iteration, the pricing problem is solved for every user and the exact minimum reduced cost trajectory $p^j_{(1)}$ is included in the columns added to the RMP whenever it is improving. This is the forced-optimal-column rule. Therefore, whenever an improving pricing column exists for any user, at least one new column is available to be added to the RMP. Since the complete trajectory set is finite, only finitely many such additions can occur.

When no improving trajectory exists for any user, the minimum reduced cost satisfies $\bar c_{p^j_{(1)}} \geq 0, \; \forall j \in \mathcal M$. This is the standard column-generation optimality condition, and hence the current RMP is optimal for the full LP relaxation of the node. Therefore its objective value equals $z^\star_{\mathrm{LP}}$.

Finally, when the B\&B search is allowed to run to completion, each node is evaluated using its LP relaxation and branching is applied until all remaining nodes are either infeasible, fathomed by their bounds, or integral. Thus, the standard B\&P argument yields a globally optimal solution of Problem \eqref{integer_master_problem}.
\end{IEEEproof}

\section{Online Handover-Aware Visibility-Duration Policy} \label{online_policy}
We next consider an online setting in which future user demands are unavailable. Although future capacity feasibility cannot be predicted accurately, satellite visibility windows, current network loads, satellite capacities, and handover costs remain available in real time. Based on these quantities, we propose a low-complexity, real-time deployable handover-aware visibility-duration (HAVD) policy.

The scoring function is designed around three competing objectives:
\begin{enumerate}
  \item \textit{Visibility persistence ($\alpha_1$):} Prefer satellites that will remain geometrically visible to $j$ for longer, reducing the likelihood of a forced handover in the near future.
  \item \textit{Handover aversion ($\alpha_2$):} Penalize switching away from the current serving satellite; this is the primary objective driver and so $\alpha_2 > \alpha_1, \alpha_2 > \alpha_3$.
  \item \textit{Congestion avoidance ($\alpha_3$):} Penalize satellites that are already heavily loaded by users processed earlier in the same slot, to spread load and reduce future capacity pressure.
\end{enumerate}

Consider a user $j \in \mathcal{M}$ currently associated with satellite $i$ at time slot $t$. Let $\mathcal{V}_t^j$ denote the set of satellites visible to user $j$. For each candidate satellite $i' \in \mathcal{V}_t^j$, the remaining visibility duration is defined as,
\begin{equation*}
  \tau_{\mathcal{V}}(i') = \max\bigl\{\Delta \geq 0 : i' \in \mathcal{V}^j_{t'},\;
  \forall t' \in \{t,\ldots,t+\Delta\}\bigr\}.
\end{equation*}
The quantity $\tau_{\mathcal{V}}(i')$ is the length of the contiguous future window over which $i'$ remains visible to $j$, evaluated purely from orbital geometry. Let
\begin{equation*}
    CL^{i'}_t = \sum_{k \in \mathcal{M}} \frac{D_t^k}{R_{t}^{i',k}} X_{t}^{i',k}
\end{equation*}
be the current load of satellite $i'$. Before scoring, every candidate $i' \in \mathcal{V}^j_t$ is tested:
\begin{equation}\label{eq:gate}
  \mathrm{CL}^{i'}_t + \frac{D_t^j}{R_{t}^{i',j}} > C^{i'} + \varepsilon_{small} \;\Rightarrow\; \text{skip } i',
\end{equation}
where $\varepsilon_{small}$ is a small constant. Candidates failing \eqref{eq:gate} are excluded from scoring entirely. This hard gate is the primary mechanism guaranteeing capacity feasibility. If no visible satellite satisfies the capacity gate, the policy declares a service outage for that slot. Since satellites may possess heterogeneous capacities, the corresponding utilization is defined as,
\begin{equation*} 
    U_f(i',t) = \frac{CL^{i'}_t}{C^{i'}}.
\end{equation*}
Now, consider, a normalized handover penalty as below:
\begin{equation*}
  H_{\mathrm{norm}}(i',j,t) = \begin{cases}
    0  & i' = \hat{i}^j_{t-1}, \\
    \dfrac{H^{\hat{i}^j_{t-1},\,i'}}{H_{\max}} & \text{otherwise},
  \end{cases}
\end{equation*}
where $\hat{i}^j_{t-1}$ is user $j$'s satellite at slot $t-1$ and $H_{\max} = \max_{i,i'} H^{i,i'}$.

The proposed HAVD policy assigns the following score to each visible satellite:
\begin{align}
  \Psi_{\mathrm{HAVD}}&(i') \notag \\
  &= \alpha_1 \tau_{\mathcal{V}, \mathrm{norm}}(i') - \alpha_2 H_{\mathrm{norm}}(i',j,t) - \alpha_3 U_f(i',t),\label{eq:phi_havd}
\end{align}
where $\alpha_1$, $\alpha_2$, and $\alpha_3$ are positive weighting coefficients, and $\tau_{\mathcal{V},\mathrm{norm}}(i') = \tau_{\mathcal{V}}(i')/T$. The selected satellite is given by,
\begin{equation*}
i^*_{j,t} = \arg\max_{i'\in \mathcal{V}_t^j} \Psi_{\mathrm{HAVD}}(i').
\end{equation*}

The visibility-duration term promotes long future service continuity and indirectly reduces the likelihood of future handovers. The handover-cost term penalizes costly switching operations, while the utilization term discourages associations with highly loaded satellites and improves load balancing across the constellation.


\begin{proposition} \label{prop:havd_comp}
HAVD runs in $O(M \mathcal{V}_{max} T^2)$ time.
\end{proposition}
\begin{IEEEproof}
Initially, HAVD precomputes set of visible satellites as a hash set for every $(j,t)$ pair, converting the visibility list into an $O(1)$-membership structure. This preprocessing costs $O(M T \mathcal{V}_{max})$ time and $O(M T \mathcal{V}_{max})$ space.

In the main loop, at each slot $t\in\{1,\ldots,T-1\}$, for each user $j\in\mathcal{M}$, HAVD evaluates its score for every candidate satellite in the current candidate set, whose size is at most $\mathcal{V}_{max}$. The utilization and handover-penalty terms are $O(1)$ given the precomputed committed load and $H$ arrays. The visibility duration is computed by scanning forward from $t$ until the first slot at which the candidate satellite is no longer visible. Each membership test costs $O(1)$ via set of visible satellites, and hence this scan costs $O(T)$ in the worst case.

Thus, evaluating all candidates for one $(j,t)$ pair costs $O(\mathcal{V}_{max}T)$. Summing over $M$ users and $T$ time slots gives $O(M\mathcal{V}_{max}T^2)$. This dominates the preprocessing cost $O(MT\mathcal{V}_{max})$, yielding the stated complexity.
\end{IEEEproof}

\section{Offline Congestion Aware Predictive (CAP) Policy} \label{offline_policy}
HAVD makes decisions based solely on slot-$t$ information. It cannot anticipate that a currently uncongested satellite will become heavily loaded two minutes later, or that its current satellite will leave the user's visibility window shortly—potentially forcing an expensive inter-shell handover later. The CAP policy is motivated by the observation that with accurate orbit propagation, the full visibility sequence $\{\mathcal{V}^j_{t'}\}_{t'=t}^{T-1}$ and link-rate matrix $\{R^{i,j}_{t'}\}$ are available in advance.

Consider a user $j \in \mathcal{M}$ currently associated with satellite $i$ at time slot $t$. Let $\mathcal{V}_t^j$ denote the set of satellites visible to user $j$. For each candidate satellite $i' \in \mathcal{V}_t^j$, the remaining feasible service duration is defined as,
\begin{align}
  \tau&(i') \notag \\
  &= \max \bigl\{ \Delta \in \{0, \ldots, T-1\-t \} : i' \in \mathcal{V}_{t'}^j, \widehat{\mathrm{Load}}_{i'}(t') \leq C^{i'}, \notag \\
  & \qquad \forall t' \in \{t, \ldots, t + \Delta\}\bigr\}, \label{eq:tau}
\end{align}
where $\widehat{\mathrm{Load}}_{i'}(t')$ is given below:
\begin{equation*}
    \widehat{\mathrm{Load}}_{i'}(t') = \sum_{k \in \mathcal{M}} \frac{D_{t'}^k}{R_{t'}^{i',k}} X_{t'}^{i',k}, \quad t' > t,
\end{equation*}
where user $j$ is counted as assigned to $i'$ for the purpose of this computation. Thus $\tau(i')$ is the length of the maximal contiguous feasible window starting at $t$.  It is zero if $i'$ is already infeasible at slot $t$. Then, capacity slack at $t'$ is defined as,
\begin{equation*}
  S_{i'}(t') = C^{i'} - \widehat{\mathrm{Load}}_{i'}(t').
\end{equation*}
The minimum congestion slack can be expressed as:
\begin{equation}\label{eq:ms}
  \mathrm{MS}(i') = \min_{t \leq t' \leq t + \tau(i')} S_{i'}(t').
\end{equation}
$\mathrm{MS}$ captures the tightest congestion margin over the entire service window; a large value implies robustness to demand variability.

The proposed CAP policy assigns the following score to each visible satellite:
\begin{equation}\label{eq:phi_cap}
  \Phi_{\mathrm{CAP}}(i') = w_1\,\frac{\tau(i')}{T-t}
    + w_2\,\frac{\mathrm{MS}(i')}{C^{i'}} - w_3\,\frac{H^{\hat{i}^j_{t-1},\,i'}}{H_{\max}}.
\end{equation}
Here, $\tau(i')/(T-t)$ is the fraction of the remaining horizon for which $i'$ is predicted to be feasible, and $\text{MS}(i')/C^{i'}$ is the normalized minimum slack.

\begin{proposition} \label{prop:cap_comp}
CAP runs in $O\!\left(M^2 \mathcal{V}_{max} T^2\right)$ time.
\end{proposition}
\begin{IEEEproof}
The CAP policy operates in an offline setting where future demands $D^j_{t'}$, data rates $R^{i,j}_{t'}$, satellite capacities $C^i$, and visibility windows are fully known over the entire horizon $\mathcal{T}$. At each slot $t$ and for each user $j$, every candidate satellite $i' \in \mathcal{V}^j_t$ is evaluated by computing the feasible service duration $\tau(i')$ \eqref{eq:tau} and the minimum congestion margin $MS(i')$ \eqref{eq:ms} over the window $[t,\, t + \tau(i')]$. Both quantities require scanning the future load $\mathrm{Load}_{i'}(t')$ for $t' \in \{t, \ldots, T-1\}$, which involves aggregating, for each future slot $t'$, the load coefficients $D^k_{t'}/R^{i',k}_{t'}$ over all users $k \in \mathcal{M}$ predicted to reside on satellite $i'$ at slot $t'$. This aggregation costs $O(M T)$ per candidate, so evaluating all candidates for a single $(t, j)$ pair costs $O\!\left(M \mathcal{V}_{max} T\right)$. Summing over all $M$ users and all $T$ slots, the total complexity of the CAP policy is $O\!\left(M^2 \mathcal{V}_{max} T^2\right)$. The quadratic dependence on both $M$ and $T$ reflects the cost of evaluating full-horizon future loads for every user--candidate pair at every slot, which is feasible in an offline setting but precludes real-time deployment.
\end{IEEEproof}

\section{Local-Search Repair Under Demand Perturbation} \label{sec:local_search_repair}
B\&P and MCS-B\&P obtain an optimal or near-optimal assignment $\hat{x}$ for a demand matrix $D \in \mathbb{R}_{\geq 0}^{M\times T}$. Let $\hat{x}_{j,t}\in\mathcal{N}$ denote the satellite assigned to user $j$ at slot $t$, where $\mathcal{N}$ is the set of satellites. In practice, realized demand may deviate from $D$ due to forecast errors and traffic fluctuations. Re-solving B\&P for each demand realization is computationally expensive. We therefore develop a local-search repair method, initialized with $\hat{x}$, that adapts the assignment to perturbed demand using two complementary move types to balance solution quality and computational cost \cite[Ch.~12]{kleinberg2006algorithm}.

\subsection{Problem Formulation} \label{sssec:ls_formulation}
Let $\delta_s \in [-1,1]$ denote the demand perturbation magnitude. The perturbed demand is
\begin{equation*}
    \tilde{D}_t^j = Y_t^j D_t^j,\qquad  Y_t^j \sim \mathrm{Uniform}[1-|\delta_s|,\,1+|\delta_s|].
\end{equation*}
Since $R_t^{i,j}$ is unchanged, the perturbed load coefficient is
\begin{equation*}
    \tilde{l}_t^{i,j} =
    \begin{cases}
        \tilde{D}_t^j/R_t^{i,j}, & R_t^{i,j}>0,\\
        \infty, & R_t^{i,j}=0.
    \end{cases}
\end{equation*}
For an assignment $x$, the load on satellite $i$ at slot $t$ is
\begin{equation*} 
    L_{i,t}(x) = \sum_{j:x_{j,t}=i}\tilde{l}_t^{i,j}, \quad i\in\mathcal{N}, \quad t=0,\ldots,T-1.
\end{equation*}
The handover cost and total capacity violation are
\begin{align*}
    \mathrm{cost}(x) &= \sum_{j=1}^{M}\sum_{t=1}^{T-1} H^{x_{j,t-1},x_{j,t}},\\
    \mathrm{viol}(x) &= \sum_{i\in\mathcal{N}}\sum_{t=0}^{T-1} \max\{0,L_{i,t}(x)-C^i\}. 
\end{align*}
We minimize the penalized objective
\begin{equation} \label{eq:ls_phi}
    \Xi(x)=\mathrm{cost}(x) + \omega_{\mathrm{pen}}\mathrm{viol}(x),
\end{equation}
where $\omega_{\mathrm{pen}}>0$ penalizes capacity violations. The search is initialized at $x^{(0)}=\hat{x}$ and seeks to reduce $\Xi(x)$ under the perturbed demand $\tilde{D}$ without resolving B\&P.

\subsection{Neighborhood Structures} \label{sssec:ls_neighborhoods}
We consider two complementary neighborhoods: single-slot reassignment and full-trajectory re-optimization for one user.

\paragraph{Single-flip neighborhood $\mathcal{Z}_1$} \label{def:n1}
For an assignment $x$, $\mathcal{Z}_1(x)$ contains assignments $x'$ differing from $x$ at exactly one coordinate $(j,t)$, where $x'_{j,t}=i_{\mathrm{new}} \in \mathcal{V}_t^j \setminus \{ x_{j,t} \}$ and $\mathcal{V}_t^j$ is the set of satellites visible to user $j$ at slot $t$ with positive achievable rate. Recomputing the objective from scratch costs $O(MT)$ per candidate. Instead, incremental updates evaluate each candidate in $O(1)$ time, since a single flip affects at most two handover edges and two satellite--slot loads. Let $i_{\mathrm{old}} = x_{j,t}$, $i_{\mathrm{prev}} = x_{j,t-1}$, and $i_{\mathrm{next}} = x_{j,t+1}$. The handover-cost change is
\begin{align}
    \Delta \mathrm{cost} =& \big[ H^{i_{\mathrm{prev}}, i_{\mathrm{new}}} - H^{i_{\mathrm{prev}}, i_{\mathrm{old}}} \big] \notag\\ 
    &+ \big[ H^{i_{\mathrm{new}}, i_{\mathrm{next}}} - H^{i_{\mathrm{old}}, i_{\mathrm{next}}} \big], \label{eq:ls_delta_cost}
\end{align}
where terms involving unavailable neighboring slots are omitted. The capacity-violation change is
\begin{align}
    \Delta \mathrm{viol} =& \big[ g(L_{i_{\mathrm{old}},t} - \tilde{l}_t^{i_{\mathrm{old}},j}) - g(L_{i_{\mathrm{old}},t}) \big] \notag \\
    &+ \big[ g(L_{i_{\mathrm{new}},t} + \tilde{l}_t^{i_{\mathrm{new}},j}) - g( L_{i_{\mathrm{new}},t}) \big], \label{eq:ls_delta_viol}
\end{align}
where $g(L_{i,t})=\max(0,L_{i,t}-C^i)$. Thus, evaluating all candidate satellites for a fixed $(j,t)$ takes $O(\mathcal{V}_{\max})$, where $\mathcal{V}_{\max} = \max_{j,t}|\mathcal{V}_t^j|$, compared with $O(M T \mathcal{V}_{\max})$ when each candidate is evaluated from scratch.

\paragraph{Trajectory re-optimization neighborhood $\mathcal{Z}_2$} \label{def:n2}
For user $j$, $\mathcal{Z}_2^{(j)}(x)$ contains assignments that keep all other users unchanged while allowing user $j$'s entire $T$-slot trajectory to vary over the feasible satellites. The overall neighborhood is
\begin{equation*}
    \mathcal{Z}_2(x) = \bigcup_{j=1}^{M}\mathcal{Z}_2^{(j)}(x).
\end{equation*}
While $\mathcal{Z}_1$ provides inexpensive local refinement, it may not capture beneficial coordinated changes across multiple slots. The $\mathcal{Z}_2$ neighborhood addresses this limitation by re-optimizing one user's complete trajectory while fixing the others. Although it may contain exponentially many trajectories, dynamic programming finds the best trajectory conditional on the other users efficiently. This richer neighborhood is motivated by the importance of neighborhood design in local-search \cite[Ch. 12.5]{kleinberg2006algorithm}.

\subsection{Algorithm} \label{sssec:ls_algorithm}
The repair procedure consists of at most $\Lambda$ rounds, each comprising a complete $\mathcal{Z}_2$ sweep followed by two $\mathcal{Z}_1$ descent phases: feasibility repair and handover-cost refinement. The procedure terminates when a complete round yields no improvement, the maximum number of rounds is reached, or the wall-clock deadline $\tau$ is reached.

\paragraph{Trajectory re-optimization using $\mathcal{Z}_2$.}
For each user $j$, all other users' trajectories are fixed. Let $x_{-j}$ denote their assignments. The neighborhood is
\begin{equation*}
    \mathcal{Z}_2^{(j)}(x) = \left\{ x' : x'_{k,t}=x_{k,t},\ \forall k \neq j,\     x'_{j,t} \in \mathcal{V}_t^j, \ \forall t \right\}.
\end{equation*}
Define the modified load excluding user $j$ as
\begin{equation*}
    L^{-j}_{i,t} = L_{i,t}(x) - \mathbf{1}_{\{ x_{j,t} = i \}} \tilde{l}^{i,j}_t.
\end{equation*}
If user $j$ is assigned to satellite $i$ at slot $t$, the resulting load is $L^{-j}_{i,t} + \tilde{l}^{i,j}_t$. The penalized objective is \eqref{eq:ls_phi}. Since the other users are fixed, their contribution is constant, and the part of the objective affected by user $j$ is
\begin{align*}
    \Xi_j (x'_{j,\cdot} \mid x_{-j}) = & \sum_{t=1}^{T-1} H^{x'_{j,t-1},x'_{j,t}} \notag\\
    & + \omega_{\mathrm{pen}} \sum_{t=0}^{T-1} g\!\left( L^{-j}_{x'_{j,t},t} + \tilde{l}^{x'_{j,t},j}_t \right),
\end{align*}
where $g(L_{i,t})=\max(0,L_{i,t}-C^i)$. Let $\Theta_t(i)$ be the minimum accumulated objective over slots $0, \ldots,t$ among feasible trajectories ending at satellite $i$ at slot $t$. Initialize
\begin{align*}
    \Theta_0(i) =&\ \omega_{\mathrm{pen}}\, g \! \left(L^{-j}_{i,0} + \tilde{l}^{i,j}_0 \right), && i \in \mathcal{V}_0^j.
\end{align*}
For $t = 1, \ldots, T-1$ and each $i \in \mathcal{V}_t^j$, compute
\begin{align*}
    \Theta_t(i) =& \min_{i' \in \mathcal{V}_{t-1}^j} \left[ \Theta_{t-1}(i') + H^{i,i'} \right] \notag\\
    & + \omega_{\mathrm{pen}} g \! \left(L^{-j}_{i,t} + \tilde{l}^{i,j}_t \right).
\end{align*}
During this recursion, store the minimizing predecessor
\begin{equation*}
    P_t(i) = \arg\min_{i' \in \mathcal{V}_{t-1}^j} \left[ \Theta_{t-1}(i') + H^{i,i'}\right].
\end{equation*}
Once the recursion reaches stage $T-1$, select the optimal final satellite
\begin{equation*}
    i^*_{T-1} = \arg\min_{i \in \mathcal{V}_{T-1}^j} \Theta_{T-1}(i),
\end{equation*}
and backtrack using $i^*_{t-1} = P_t(i^*_t)$ for $t = T-1, \ldots, 1$ to obtain the trajectory $x^*_{j,\cdot} = (i^*_0,i^*_1, \ldots, i^*_{T-1})$. Construct the candidate assignment $x' = (x_{-j}, x^*_{j, \cdot})$. Accept $x'$ if $\Xi(x') < \Xi(x)$; otherwise, retain the current assignment. Process users sequentially for $j = 1, \ldots, M$, updating the assignment and load matrix after each accepted move. The $\mathcal{Z}_2$ sweep is complete once all users have been processed. If the DP is interrupted by the wall-clock deadline $\tau$ before reaching stage $T-1$, discard its partial results and retain the user's previous trajectory.

\paragraph{Feasibility repair using $\mathcal{Z}_1$.}
After the complete $\mathcal{Z}_2$ sweep, identify the overloaded satellite--slot cells:
\begin{equation*}
    \text{Overload}(x)= \{(i,t):L_{i,t}(x)>C^i\}.
\end{equation*}
Consider single-slot flips for users assigned to overloaded cells. At each iteration, select the eligible flip yielding the largest strict decrease in capacity violation, i.e., the most negative $\Delta \mathrm{viol}< 0$. Update the assignment and load matrix after each accepted flip, and repeat until no eligible flip strictly reduces $\mathrm{viol}(x)$ or the deadline $\tau$ is reached.

\paragraph{Handover-cost refinement using $\mathcal{Z}_1$.}
After feasibility repair, consider single-slot flips satisfying $\Delta \mathrm{cost} < 0$, $\Delta \mathrm{viol} \leq 0$. Accept an eligible flip, update the assignment and load matrix, and repeat until no such improving flip remains or the deadline $\tau$ is reached. This phase reduces handover cost without increasing capacity violation.

\begin{theorem}[Complexity] \label{thm:ls_complexity}
A complete $\mathcal{Z}_1$ sweep over all $(j,t)$ coordinates requires $O(M \mathcal{V}_{\max} T)$ time, while a complete $\mathcal{Z}_2$ pass over all users requires $O(M \mathcal{V}_{\max}^2 T)$ time. If each of
the two $\mathcal{Z}_1$ phases performs at most $S$ passes per round, the local-search repair procedure has worst-case running time
\begin{equation*} 
    O\!\left( \Lambda M \mathcal{V}_{\max} T (\mathcal{V}_{\max} + S) \right).
\end{equation*}
\end{theorem}

\begin{IEEEproof}
For a single $\mathcal{Z}_1$ move, only two handover edges and two satellite--slot loads are affected. Thus, each candidate satellite can be evaluated in $O(1)$ time using \eqref{eq:ls_delta_cost} and \eqref{eq:ls_delta_viol}. Since at most $\mathcal{V}_{\max}$ candidate satellites are considered at each of
the $MT$ coordinates, one complete $\mathcal{Z}_1$ pass requires $O(M\mathcal{V}_{\max}T)$ time.

For a single $\mathcal{Z}_2$ move, the DP evaluates transitions between at most $\mathcal{V}_{\max}$ feasible satellites at consecutive time slots, requiring $O(\mathcal{V}_{\max}^2)$ time per slot. Therefore, re-optimizing one user's trajectory requires $O(T \mathcal{V}_{\max}^2)$ time, and a complete pass over all $M$
users requires $O(M T \mathcal{V}_{\max}^2)$ time. Across at most $\Lambda$ rounds, the $\mathcal{Z}_2$ passes contribute $O(\Lambda M T \mathcal{V}_{\max}^2)$ time.

Each of the two $\mathcal{Z}_1$ phases performs at most $S$ passes per round, contributing $O(\Lambda S M T \mathcal{V}_{\max})$ time in total. Summing the contributions of all three phases gives
\begin{equation*}
\resizebox{252pt}{!}{$
    O \! \left( \Lambda M T \mathcal{V}_{\max}^2 + \Lambda S M T \mathcal{V}_{\max} \right) = O \! \left( \Lambda M \mathcal{V}_{\max} T (\mathcal{V}_{\max} + S) \right), $}
\end{equation*}
which establishes the result.
\end{IEEEproof}

For fixed $\Lambda$ and $S$, the repair procedure has polynomial running time in $M$, $T$, and $\mathcal{V}_{\max}$. Moreover, because $\mathcal{V}_{\max} \ll N$ in the considered LEO setting, the repair procedure explores a restricted set of rate-feasible satellite assignments rather than re-solving the full B\&P problem. This makes local-search repair a computationally attractive alternative for adapting a precomputed assignment to perturbed demand. Its runtime advantage over repeated B\&P re-optimization is evaluated empirically in Section \ref{results}.

\section{Results} \label{results}
We evaluate the proposed RL-based MCS-B\&P against Classical B\&P, HAVD, and CAP using a realistic Gen-1 Starlink LEO constellation. Experiments were conducted on an Exatron GPU node of the Prajna HPC cluster at IIT Bombay, with 16 CPU cores and two NVIDIA A40 GPUs. The simulation code is available at \cite{nalavade2026GitHub}.

\begin{figure*}[ht] 
    \centering
    \includegraphics[width=0.6\textwidth]{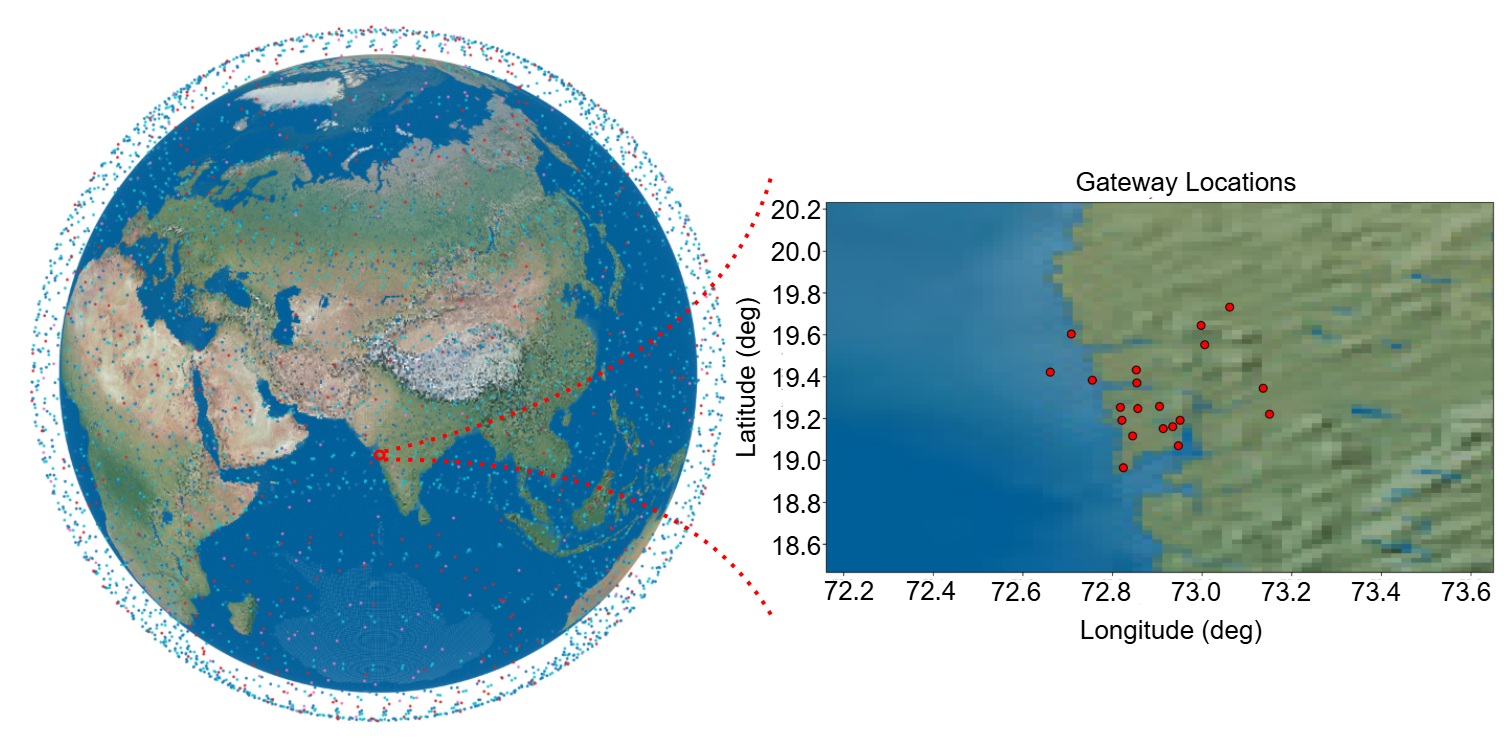} 
    \caption{The figure provides a visualization of the Starlink constellation at $t=0$ and $30$ gateway locations for demonstration purpose.}
    \label{fig:constellation}
\end{figure*}

The constellation comprises three orbital shells, as summarized in Table \ref{tab:shells}. Satellite positions are propagated using Keplerian orbital mechanics at a slot duration of $\delta_t=2$s. A satellite is visible to a user if its elevation angle exceeds $10^\circ$. We randomly select $M$ gateways from Mumbai, Thane, and Palghar, Maharashtra, India. Scenarios in which the selected gateways cannot be covered by a common satellite are treated as independent
problems. Downlink rates $R_t^{i,j}$ are computed using the Shannon formula and the link-budget parameters in Table \ref{tab:parameter_values}, based on \cite{FCC_SES_02626_2023}. Fig. \ref{fig:constellation} illustrates the Starlink constellation at time $t=0$, where the satellites orbiting the Earth are depicted as small dots, and the collection of ground users is represented by red circles.

\begin{table}[!h]
\centering
\caption{Downlink Parameters for NGSO Satellite Link Budget \cite{FCC_SES_02626_2023}.}
 \label{tab:parameter_values}
\begin{tabular}{ccccc}
\toprule
Parameters & Values & & Parameters & Values \\
\midrule
Frequency & 19 GHz & & Noise Figure & 1.5 dB \\
Bandwidth & 250 MHz & & Misc. Loss & 2 dB \\
EIRP & 70 dBm & & Rain Fade & 2 dB \\
Receiver Gain & 47 dBi & & Elevation Loss Factor & 0.02 \\
\bottomrule
\end{tabular}
\end{table}

\begin{table}[!h]
\centering
\caption{Starlink Gen-1 Shell Parameters}
\label{tab:shells}
\begin{tabular}{ccccc}
\toprule
Shell & Alt.\ (km) & Inc.\ ($^\circ$) & Planes & Satellites \\
\midrule
0 & 550 & 53.0 & 72 & 1584 \\
1 & 570 & 70.0 & 36 & 360  \\
2 & 560 & 97.6 & 6  & 348  \\
\bottomrule
\end{tabular}
\end{table}

The handover cost matrix assigns increasing cost ranges to intra-plane, intra-shell cross-plane, and inter-shell transitions expressed as below:
\begin{equation*}
\resizebox{252pt}{!}{$
  H_{i,i'} = \begin{cases}
    0 & i = i', \\
    \mathrm{Uniform}\ [0.5,\, 1.0] & \text{intra-plane } (i \neq i'), \\
    \mathrm{Uniform}\ [1.5,\, 3.0] & \text{intra-shell, cross-plane}, \\
    \mathrm{Uniform}\ [4.0,\, 8.0] + 0.5\,|s_i - s_{i'}| & \textit{inter-shell},
  \end{cases}$}
\end{equation*}
where $\mathrm{Uniform}\ [a,b]$ denotes a uniform random draw and $|s_i - s_{i'}|$ is the shell-index gap. Satellite capacities are sampled from $\mathrm{Uniform}[5.0,6.5]$ for loosely congested traffic and $\mathrm{Uniform}[3.0,4.5]$ for highly congested traffic. The user $j$ generates demand $D^j_t \sim \text{Exponential}(\bar{d})$, clipped to $[0, D_{\max}]$, with a probability $p_{idle}$ of a zero-demand (idle) slot:
\begin{equation*}
  D^j_t = \begin{cases} 0 & \text{w.p. } p_{idle}, \\ \min(\tilde{D}, D_{\max})
  & \text{w.p. } 1-p_{idle}, \end{cases}
  \quad \tilde{D} \sim \text{Exp}(\bar{d}).
\end{equation*}
The following parameters are used to generate the demand: $\bar{d} = 95$\,Mbps, $D_{\max} = 400$\,Mbps, $p_{idle} = 0.02$.

Table \ref{tab:results_congestion} compares the computational performance of the four methods for $T=300$ under different congestion levels and user populations. For both highly congested scenarios, Classical B\&P and MCS-B\&P are subject to a wall-clock limit of $10$ hours ($36000$s). In the loosely congested scenario with $M=100$, MCS-B\&P achieves the same optimal handover cost as Classical B\&P while reducing the solution time by approximately $10.6\%$. Under high congestion with $M=115$, both methods obtain the same handover cost, but MCS-B\&P completes the optimization in $25061.4$s, compared with the $10$-hour limit for Classical B\&P. For $M=125$, Classical B\&P fails to obtain a solution within the $10$-hour limit, whereas MCS-B\&P finds a solution with a handover cost of $193.018$ in $23269.4$s. These results demonstrate that MCS-B\&P improves computational efficiency while preserving solution quality, particularly as the problem becomes more challenging. HAVD and CAP provide substantially shorter solution times but incur considerably higher handover costs across all three scenarios.

\begin{table}[!h]
\centering
\caption{Performance comparison under different congestion levels.}
\label{tab:results_congestion}
\begin{tabular}{lccc}
\toprule
Method & Handover Cost & Time (s) & Gap (\%) \\
\midrule
\multicolumn{4}{c}{\textit{Loosely congested scenario with $M=100$}} \\
\midrule
Classical B\&P & 53.2624 & 3237.5 & 0 \\
MCS-B\&P       & 53.2624 & 2893.9 & 0 \\
HAVD           & 139.8105 & 7.4 & 162.49 \\
CAP            & 134.8800 & 74.4 & 153.24 \\
\midrule
\midrule
\multicolumn{4}{c}{\textit{Highly congested scenario with $M=125$}} \\
\midrule
Classical B\&P & - & 36000.0 & - \\
MCS-B\&P       & 193.018 & 23269.4 & 0 \\
HAVD           & 510.9527 & 5.6 & 164.72 \\
CAP            & 495.3736 & 67.9 & 156.65 \\
\midrule
\midrule
\multicolumn{4}{c}{\textit{Highly congested scenario with $M=115$}} \\
\midrule
Classical B\&P & 167.5589 & 36000.0 & 0 \\
MCS-B\&P       & 167.5589 & 25061.4 & 0 \\
HAVD           & 461.4749 & 5.1 & 175.41 \\
CAP            & 458.3581 & 62.5 & 173.55 \\
\bottomrule
\end{tabular}
\end{table}

Table \ref{tab:results_metrics} compares the B\&B and CG effort of Classical B\&P and MCS-B\&P across all three scenarios. MCS-B\&P consistently reduces the number of processed nodes, pruned nodes, and CG iterations. For $M=100$, it reduces CG iterations from $607$ to $384$ and processed nodes from $215$ to $129$. Under high congestion, the reductions are similarly substantial: CG iterations decrease from $8073$ to $3376$ for $M=125$ and from $8964$ to $3863$ for $M=115$. These results demonstrate that the multiple-column selection strategy reduces both column-generation effort and the size of the B\&B search, thereby improving computational efficiency.

\begin{table}[!h]
\centering
\caption{B\&B and column-generation statistics for the loosely congested scenario.}
\label{tab:results_metrics}
\begin{tabular}{lccc}
\toprule
Method & Nodes Processed & Nodes Pruned & CG Iterations \\
\midrule
\multicolumn{4}{c}{\textit{Loosely congested scenario with $M=100$}} \\
\midrule
Classical B\&P & 215 & 106 & 607 \\
MCS-B\&P       & 129 & 63  & 384 \\
\midrule
\midrule
\multicolumn{4}{c}{\textit{Highly congested scenario with $M=125$}} \\
\midrule
Classical B\&P & 2973 & 1413 & 8073 \\
MCS-B\&P       & 1357 & 677  & 3376 \\
\midrule
\midrule
\multicolumn{4}{c}{\textit{Highly congested scenario with $M=115$}} \\
\midrule
Classical B\&P & 3443 & 1671 & 8964 \\
MCS-B\&P       & 1729 & 864  & 3863 \\
\bottomrule
\end{tabular}
\end{table}

\begin{figure}[!t]
\centering
\subfloat[]{\includegraphics[width=0.49\columnwidth]{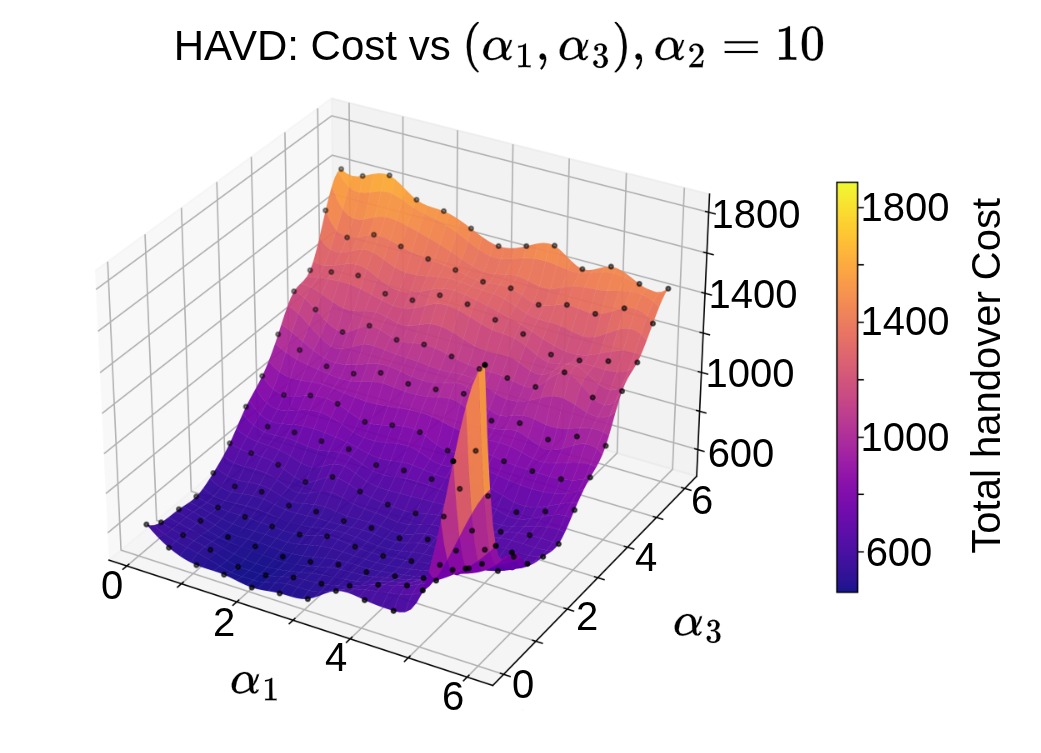}%
\label{fig:havd}}
\hfil
\subfloat[]{\includegraphics[width=0.49\columnwidth]{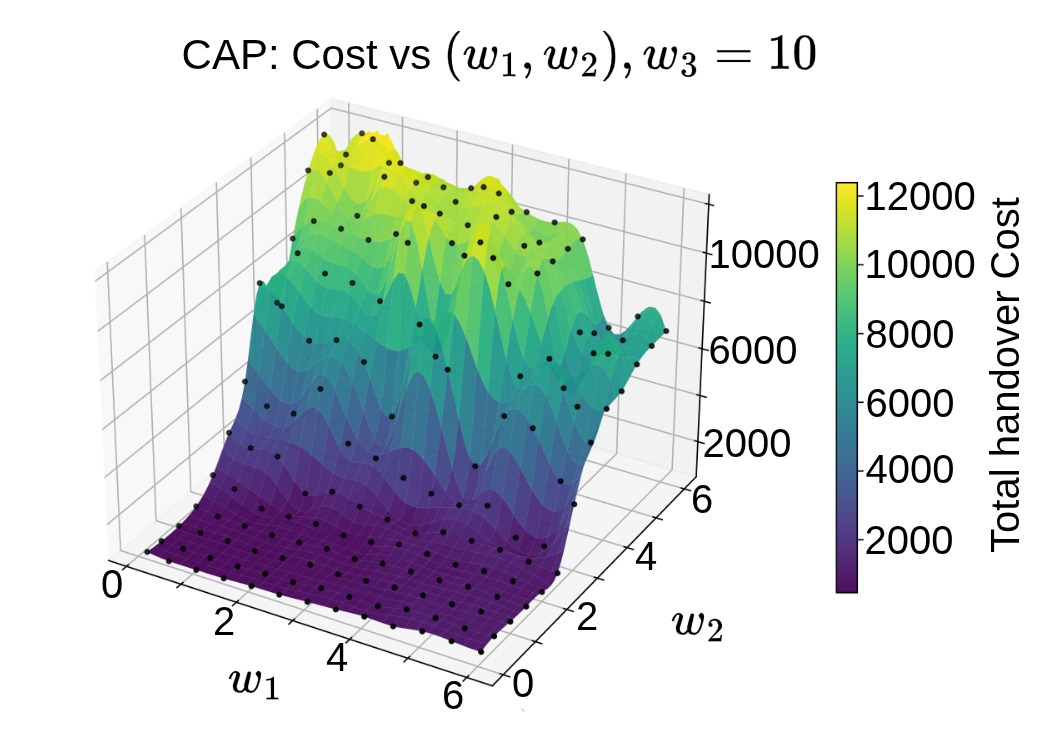}%
\label{fig:cap}}
\caption{Sensitivity of the total handover cost to the weighting parameters of the HAVD and CAP policies: (a) HAVD, varying $\alpha_1$ and $\alpha_3$ with $\alpha_2 =10$; (b) CAP, varying $w_1$ and $w_2$ with $w_3 = 10$. The handover-related weights are fixed at larger values to prioritize handover-cost minimization.}
\label{fig:heuristics}
\end{figure}

Fig. \ref{fig:heuristics} illustrates the sensitivity of the total handover cost to the weighting parameters of the HAVD and CAP policies. For HAVD, the cost is evaluated as a function of $\alpha_1$ and $\alpha_3$, with $\alpha_2 = 10$ fixed, whereas for CAP, it is evaluated as a function of $w_1$ and $w_2$, with $w_3 = 10$ fixed. In both policies, the handover-related weight ($\alpha_2$ for HAVD and $w_3$ for CAP) is chosen to be greater than the other two weights to prioritize handover-cost minimization. The figures show that the total handover cost is sensitive to the relative weights assigned to the competing objectives, with substantial cost increases in certain parameter regions. In particular, the HAVD surface exhibits a generally smoother variation in cost, apart from a localized sharp peak, whereas the CAP surface exhibits more pronounced variations and higher costs for larger values of the visibility and congestion-related weights. These results highlight the importance of appropriate weight selection to balance the competing objectives and achieve lower total handover cost.

Table \ref{tab:results_perturbation} compares local-search repair with exact re-optimization under different demand perturbations. In the loosely congested scenario with $M=100$, local-search consistently achieves handover costs close to those of Classical B\&P, with optimality gaps no greater than $1.41\%$ across perturbations of $10\%$--$80\%$. At an $80\%$ perturbation, local-search obtains a cost of $54.1271$ in only $10.69$s, compared with $53.3755$ in $1687.76$s for Classical B\&P. Similar performance is observed under high congestion: for $M=125$ and a $5\%$ perturbation, local-search achieves a gap of only $0.002\%$ while reducing the computation time from $33110.6$s to $8.36$s. For $M=125$ and $M=115$ under a $6\%$ perturbation, the gaps remain below $0.55\%$, with local-search completing in less than $8.4$s. These results demonstrate that local-search repair adapts efficiently to demand perturbations while maintaining solution quality close to that of exact re-optimization.

\begin{table}[!h]
\centering
\caption{Local-search repair performance under demand perturbations.}
\label{tab:results_perturbation}
\scalebox{0.98}{
\begin{tabular}{cccccc}
\toprule
\multirow{2}{*}{\begin{tabular}[c]{@{}c@{}}Demand\\ Perturbation\\ (\%)\end{tabular}} & \multicolumn{2}{c}{Classical B\&P} & \multicolumn{3}{c}{Local-Search} \\
    & \begin{tabular}[c]{@{}c@{}}Handover\\ Cost\end{tabular} & \begin{tabular}[c]{@{}c@{}}Time\\ (s)\end{tabular} & \begin{tabular}[c]{@{}c@{}}Handover\\ Cost\end{tabular} & \begin{tabular}[c]{@{}c@{}}Time\\ (s)\end{tabular} & \begin{tabular}[c]{@{}c@{}}Optimality\\ Gap (\%)\end{tabular} \\
    \midrule
    \multicolumn{6}{c}{\textit{Loosely congested scenario with $M=100$}} \\
    \midrule
    10  & 53.2497 & 779.37 & 53.2625 & 10.73 & 0.002 \\
    20  & 53.3375 & 871.51 & 53.3502 & 16.47 & 0.024 \\
    30  & 53.1747 & 100.68 & 53.2497 & 16.51 & 0.140 \\
    40  & 53.2751 & 829.52 & 53.3375 & 16.87 & 0.12 \\
    50  & 53.2497 & 100.17 & 53.2498 & 10.93 & 0.00 \\
    60  & 53.3375 & 569.10 & 53.8639 & 16.22 & 0.99 \\
    70  & 53.2625 & 514.53 & 53.4252 & 10.86 & 0.31 \\
    80  & 53.3755 & 1687.76 & 54.1271 & 10.69 & 1.41 \\
    \midrule
    \midrule
    \multicolumn{6}{c}{\textit{Highly congested scenario}} \\
    \midrule
    5 (M=125)  & 193.0141 & 33110.6 & 193.0180 & 8.36 & 0.002 \\
    6 (M=125) & 192.9994 & 67610.4 & 193.9430 & 8.38 & 0.489 \\
    6 (M=115) & 167.5874 & 15175.9 & 168.5071 & 7.57 & 0.549 \\
\bottomrule
\end{tabular}}
\end{table}

\section{Conclusion and Future Work} \label{conclusion}
This paper addressed handover cost minimization in large scale LEO satellite networks under demand and satellite-capacity constraints. We developed an RL-based MCS-B\&P framework that accelerates column generation while preserving exactness through forced inclusion of minimum reduced cost trajectories. We also proposed HAVD and CAP for fast handover decisions and a local-search repair procedure for adapting assignments to demand perturbations. Results showed that MCS-B\&P reduces computational effort while maintaining solution quality, whereas HAVD and CAP trade solution quality for speed. Under high congestion, MCS-B\&P obtained a solution within the wall-clock limit, and local-search repair efficiently adapted assignments to severe demand perturbations with small handover cost gaps. Future work will investigate adaptive parameter tuning, more realistic spatiotemporal demand models, and scalable online repair strategies under dynamic satellite visibility and time-varying network conditions.

\bibliographystyle{IEEEtran}  
\bibliography{references_sat_ho} 
\end{document}